\documentclass[aps,reprint,showpacs]{revtex4-1}

\usepackage{amssymb,amsmath} 
\usepackage{graphicx}

\usepackage{appendix}
\usepackage[small,bf]{caption}

\usepackage{color}
\usepackage[usenames,dvipsnames,svgnames,table]{xcolor}

\begin{document}

\author{Andrew J. Dunleavy }
\email{andrew.dunleavy@bristol.ac.uk}
\affiliation{Centre for Complexity Sciences, School of Chemistry, University of Bristol, Bristol BS8 1TW, United Kingdom}

\author{Karoline Wiesner}
\affiliation{School of Mathematics, Centre for Complexity Sciences, University of Bristol, Bristol BS8 1TW, United Kingdom}

\author{C. Patrick Royall}
\affiliation{School of Chemistry, University of Bristol, Bristol BS8 1TW, United Kingdom}

\title{Using mutual information to measure order in model glass-formers}
\date{\today}

\begin{abstract}
Whether or not there is growing static order accompanying the dynamical heterogeneity and increasing relaxation times seen in glassy systems is a matter of dispute. An obstacle to resolving this issue is that the order is expected to be amorphous and so not amenable to simple order parameters. We use mutual information to provide a general measurement of order that is sensitive to multi-particle correlations. We apply this to two glass-forming systems (2D binary mixtures of hard disks with different size ratios to give varying amounts of hexatic order) and show that there is little growth of amorphous order in the system without crystalline order. In both cases we measure the dynamical length with a four-point correlation function and find that it increases significantly faster than the static lengths in the system as density is increased. We further show that we can recover the known scaling of the dynamic correlation length in a kinetically constrained model, the 2-TLG.

\end{abstract}


\maketitle

\setlength{\parindent}{0.0in}
\setlength{\parskip}{0.1in}

\section{Introduction}

A central question in the physics of glass forming liquids is whether or not they develop static structure when supercooled (or compressed), and whether such structure plays an important role in the extreme slowing down that occurs as a system approaches the glass transition. Often, we expect the increase of relaxation time in a system to be accompanied by an increasing length-scale \cite{cavagna:2009, berthier:2011} and this should be the case if glassy phenomena are related to some sort of critical behaviour \cite{berthier:2011}. It is well established that many supercooled liquids exhibit a growing dynamic length-scale \cite{glotzer:2000} but pairwise correlation functions show little change in the structure upon cooling \cite{berthier:2010}.

A number of theoretical scenarios have been postulated to explain glassy behaviour. Some include static structure: such as clusters of locally-favoured order \cite{frank:1952,tarjus:2005}; or a mosaic of finite regions of amorphous order \cite{lubchenko:2007}. There are alternative explanations based on dynamic facilitation effects \cite{keys:2011} or a dynamical phase transition \cite{Elmatad:2010}. Neither of these scenarious require static structure. Futhermore, it is possible to produce glass-like behaviour with kinetically constrained models \cite{pan:2005,kronig:1994} that are designed to omit static structure, although these are not derived from the microscopic behaviour of actual glasses. 

So, the necessity of a growing static length-scale is questionable. There is good evidence that some change in structure (not necessarily related to a growing length scale) occurs  on dynamic slowing \cite{coslovich:2007,royall:2008,kawasaki:2007,malins:2012,tanaka:2010}. Some numerical studies indicate a growing static length-scale  \cite{kawasaki:2007,sausset:2010,malins:2012,tanaka:2010} and there are thermodynamic treatments \cite{karmakar:2012} which imply an increasing length-scale (although it is not measured directly from the real space configuration of particles). Other experimental \cite{berthier:2005,crauste2010} and numerical \cite{charbonneau:2012} work suggests that this length-scale increases only slightly. There is disagreement over whether static lengths in glassy systems grow with the dynamical length-scale or not  \cite{tanaka:2010, biroli:2008,sausset:2010,charbonneau:2012, berthier:2007}.

To determine exactly what is going on we need a technique that can measure amorphous order directly in a given system. In some systems there is a clear idea of what this order should look like and it can be measured with, for example, a bond-orientational order parameter \cite{kawasaki:2007,tanaka:2010}. The problem in these cases is that the technique is not general and systems with such clear and simple ordering may not be characteristic of glassy systems. A less specific approach is to look for geometrical motifs \cite{royall:2008} although this still requires that we limit our search to a set of predefined structures. Generalised order parameters have been suggested: local ``structural entropy'' s$^2$ \cite{tanaka:2010} does not rely on presupposed order but it may be confounded by dynamical information and it is only sensitive to pairwise correlations; the configurational entropy approach to measuring patch-correlation length \cite{kurchan:2011,sausset:2011} is certainly general although it is computationally unfeasible for the system studied here. We discuss both of these later in the text.

It would be useful to have a general method of measuring structure that does not require us to specify in advance what we are looking for and that is not blind to things that we did not expect. Information theory gives us a framework in which we can look for structure in an order-agnostic way. Using the concept of mutual information (see e.g. \cite{cover:1991}) we can quantify all of the dependencies between two multi-dimensional random variables. This enables us to develop techniques that are sensitive to higher order correlations and that do not depend on the structure in the system taking a presumed form. 

We use the mutual information between patches in a system's configuration as a general measurement of order. We measure this quantity in a model glass-forming system: a 2D binary mixture of hard-disks. The size ratio of the disk species is varied to alter the amount of hexatic order in the system. We derive a static length-scale from these mutual information measurements: it grows in tandem with the hexatic order correlation length in the hexatic system as density is increased. The mutual information length in the non-hexatically-ordered system varies little as density is changed. In both cases the growth in the dynamical length-scale significantly exceeds that of the static lengths. We also apply these methods to a kinetically-constrained model.

The paper is structured as follows.  We briefly review the concept of mutual information and discuss how it can be used to look for static and dynamic structure in super-cooled liquids. We investigate two systems which exhibit glassy behaviour: a 2D off-lattice binary hard disk mixture and a kinetically constrained model, the 2-TLG (lattice-based, and also 2D). We describe the discretisation procedure we use to obtain a symbolised representation of patches in the system's configuration and then calculate the mutual information between pairs of patches. These measurements are used to define a mutual information length of static order in the system. We investigate the behaviour of this length as the density of the system is varied, and compare the lengths to the dynamic correlation length in the system. The similarities between this approach and the patch-correlation length of \cite{sausset:2011} will be discussed.

\section{Information Theory}

Here we discuss the application of information theory to extract a structural or dynamic length-scale.
The fundamental information theoretic quantity is the Shannon entropy \cite{Shannon:1948}. The Shannon entropy of a random variable $X$ with a probability distribution $p(x)$ over a support $\mathcal{X}$ is given by: 
\begin{equation}
H(X) = - \sum_{x \in \mathcal{X}} p(x) \log_2 p(x)
\end{equation}
This quantity is larger for a uniform probability distribution over a broad support (phase space), and smaller when the support gets smaller, or the distribution more peaked. It is a measure of the uncertainty of the outcome of drawing a sample from the distribution.

When measuring structure, we are interested in how the configuration in one part of a system affects the configuration in another. We can think about this in information theoretic terms. If the configuration, $X$, in some part of the system (we call this a patch) influences that in another part, $Y$, then it will be the case that when $X$ is held constant the range of possible values of $Y$ is smaller than when $X$ can take any value. We can quantify this reduction in uncertainty by treating our configurations as random variables and taking the mutual information (see e.g. \cite{cover:1991}). 

The mutual information between two random variables measures the entropy difference between the marginal probability distribution of a variable, and its conditional distribution.
\begin{align}
I(X;Y) &= H(X) - H(X|Y) = H(Y) - H(Y|X) \\
&= H(X) + H(Y) - H(X,Y) \label{eqn:weuse}\\ 
&= \sum_{x \in \mathcal{X}, y \in \mathcal{Y}} p(x,y) \log_2 \frac{p(x,y)}{p(x)p(y)} \label{eqn:muinf}
\end{align}

The mutual information can be thought of as a distance (although not rigorously) between the true joint distribution of the two variables, and the distribution they would have if they were independent. In terms of configurations, the mutual information will be zero when $X$ has no influence on $Y$; it will be positive and increasing as $X$ becomes more influential; and it will take its maximum value $I(X;Y) = H(Y)$ if $Y$ is completely determined by $X$. The mutual information is symmetric in $X$ and $Y$.

We have a choice in the shape of our patches, $X$ and $Y$. When measuring mutual information in time-series data it is intuitive to divide the system at a nominal present time, $t$, and measure the mutual information between the output over some past period ($t-\tau \to t$) and the future output (from $t \to t +\tau$) \cite{Crutchfield:2003}. By varying $\tau$ it is possible to measure not only the amount of information the past of the system holds about the future, but also the length of time information persists in the system.

Here we are looking at spatial data: an analogous approach would be to divide the system in two and measure mutual information between configurations either side of the divide. However, this gives configuration spaces that are too large to sample. It is possible to approximate this approach (perfectly, under certain conditions) by measuring the mutual information of two abutting patches and varying their length (in the direction away from their interface) \cite{Feldman:2003}. The configuration space is smaller, but proportional to the length of the patches. Therefore, it can still be too big when the patches are made long enough to encompass long-range correlations. A computationally cheaper method is to measure the mutual information between two patches that are not abutting. Correlations at different lengths can be measured by varying the separation of the patches rather than increasing their size. 
  
The patch correlation length of \cite{sausset:2011} is based on the entropy of single patches rather than the mutual information between patches. The patches are centred on particles so each patch represents the configuration of particles within a radius, $r$ of a given particle. Two patches are said to belong to the same state if these configurations are the same (some difference is allowed for thermal vibration). The configurational entropy is calculated by comparing all of the particles in the system and taking the entropy of the distribution of states. This approach was inspired by the random first-order transition theory of glasses \cite{lubchenko:2007} which considers the system a mosaic of ordered tiles. This order ensures that the entropy increases sub-extensively with $r$ until patches start to encompass multiple tiles. As $r$ is increased beyond this the entropy becomes extensive: the patch correlation length measures this crossover. 

Alternatively, one could measure the mutual information between two separate patches in the system. If these are within a single tile the mutual information will be positive; if they do not share any tiles then the mutual information will be zero. If there is a crossover from sub-extensive to extensive regimes in the configurational entropy then this will be represented in the mutual information. 

The advantage of our mutual information approach is that we have defined our patches in such a way to probe different distances without increasing the size of the patches. As patch size increases it becomes harder to sample the patch distribution well: the method used in \cite{sausset:2011} can measure a maximum entropy of $\log N$. $N$ is the number of particles (hence patches) in the system so the maximum is reached when all $N$ patches are in unique states. In fact, the entropy should be lower than this maximum to ensure that all possible states have had a chance to be sampled. We reached this limit for small patches when measuring the configurational entropy for the binary hard-disk system whereas we were able to apply our method for measuring the mutual information between patches successfully. Our method does discard some information: it will not account for multipoint correlations that involve particles in the gap between the two patches; however, it will still detect the length over which static order extends in the system.
 
\section{Methods}

\begin{figure}[htbp]
\includegraphics[height=60mm]{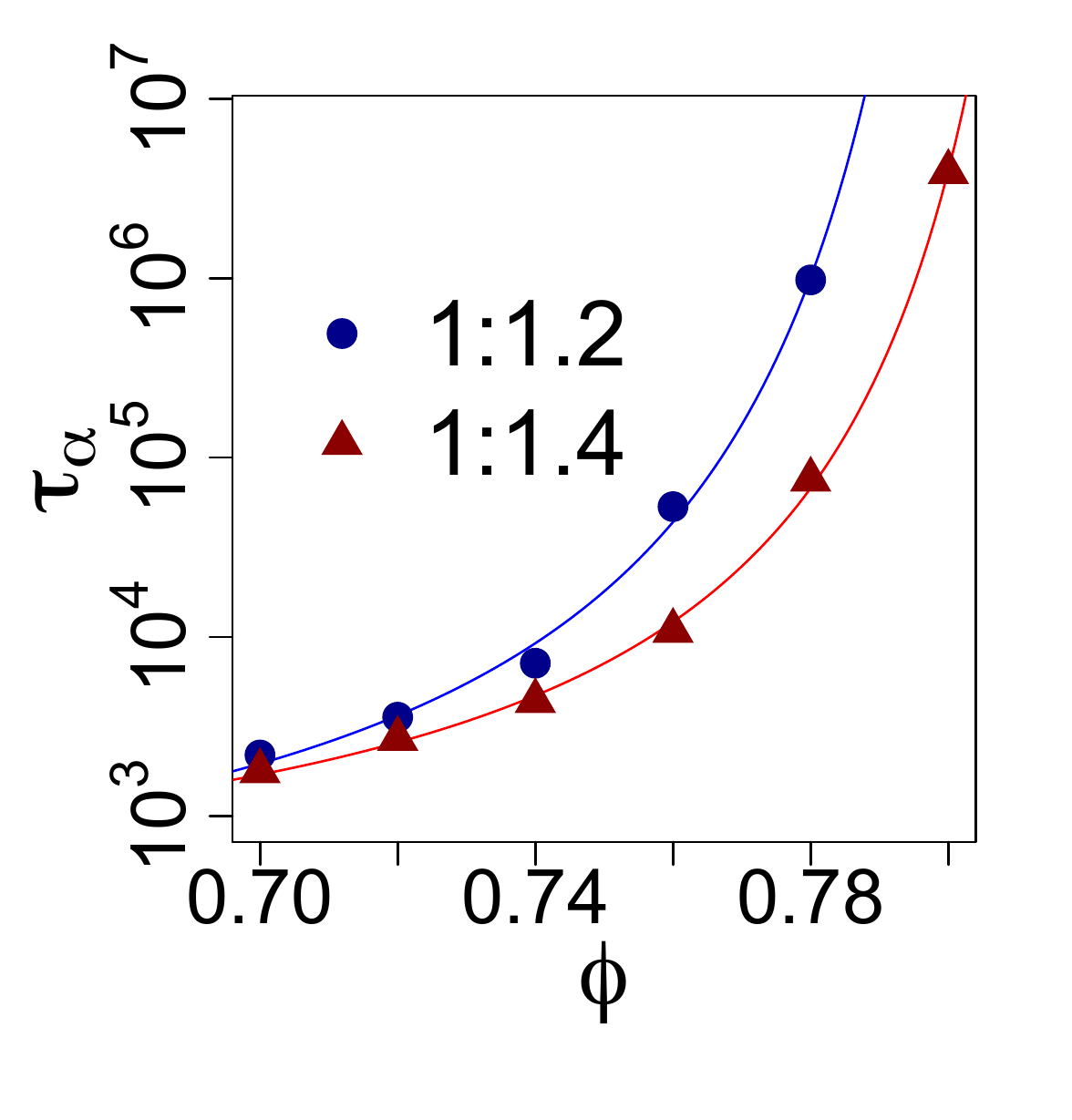}
\caption{The relaxation time $\tau_\alpha$ against area fraction $\phi$ for the two values of $R$ in the hard disk system. $\tau_\alpha$ is measured as the time taken by the self-intermediate scattering function to fall to $\exp(-1)$ \cite{hansen:2006}. The lines are VFT fits: $\tau_\alpha \propto \exp(A / (\phi_0 - \phi ))$. $\phi_0$ equals 0.82 (0.83) for the $R=1.2$ $(1.4)$ systems. }
\label{fig:angell}
\end{figure}

\begin{figure*}[htbp]
\includegraphics[height=70mm]{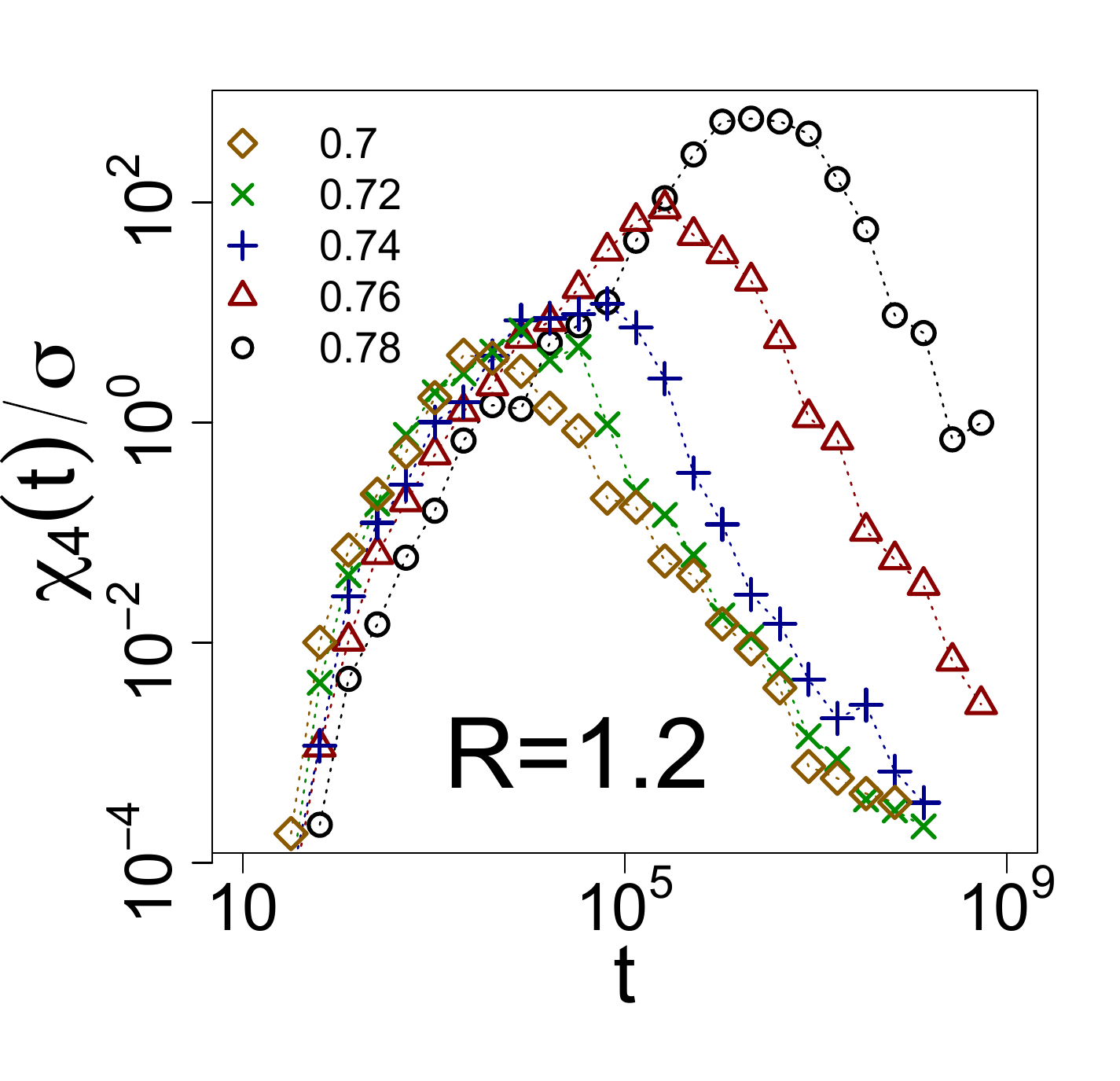}
\includegraphics[height=70mm]{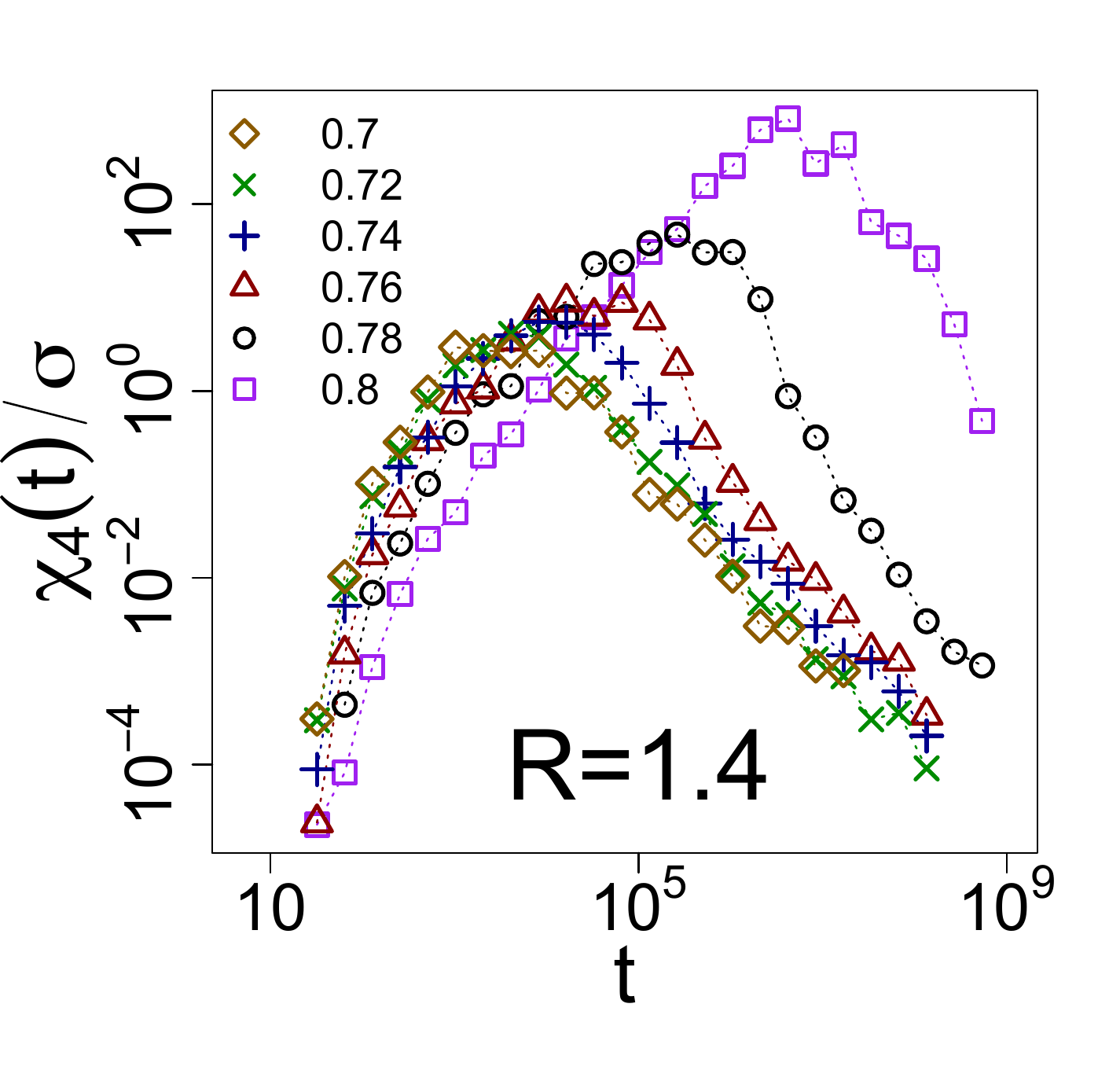}
\caption{$\chi_4(t)$ for the binary hard disk system at $R=1.2$ and $R=1.4$ . The time of the maximum $\chi_4(t=\tau_h)$ is used when calculating the length, $\xi_4$.}
\label{fig:chi4}
\end{figure*}

\begin{figure}[htbp]
\includegraphics[height=60mm]{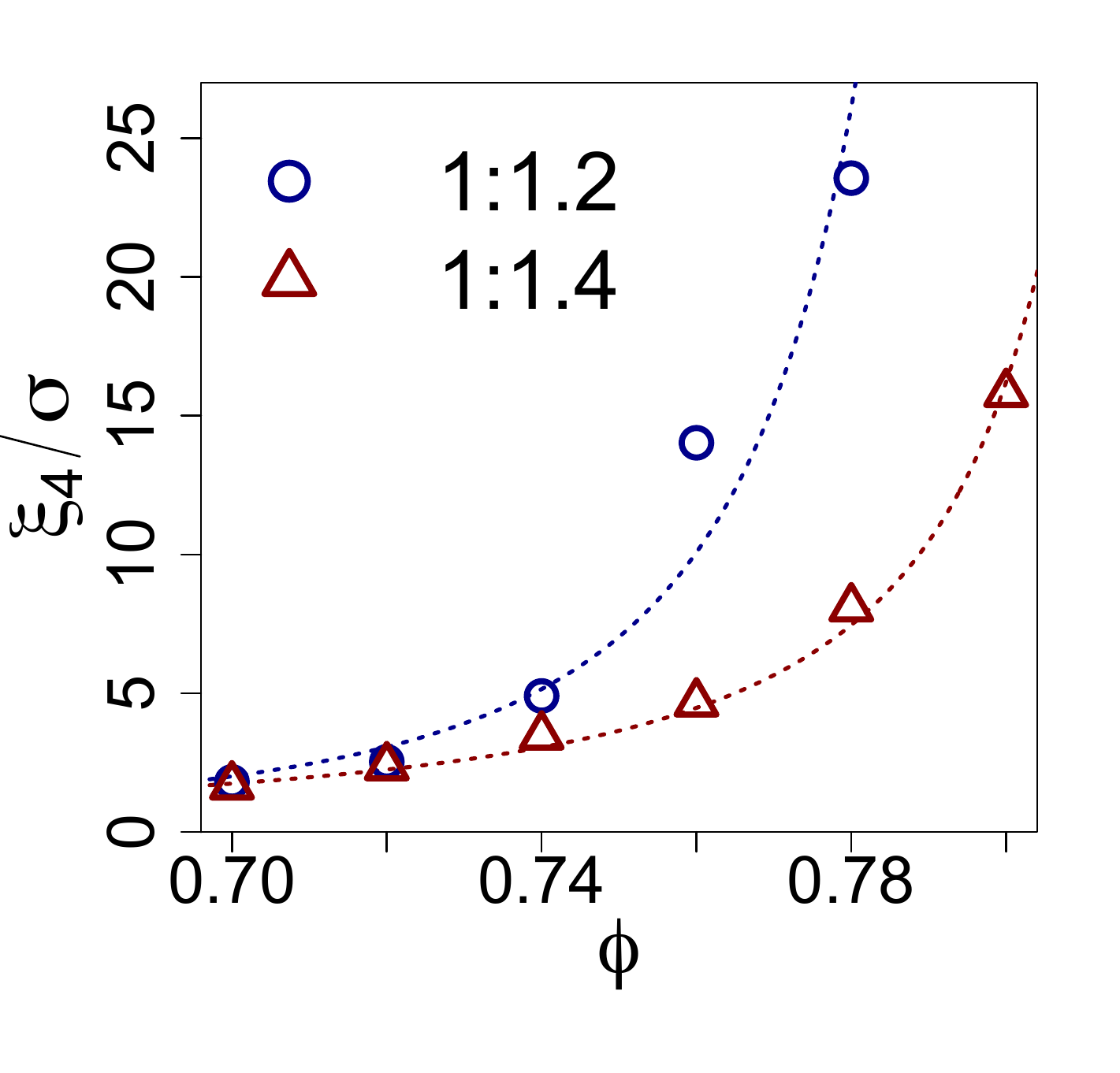}
\caption{The dynamic correlation length $\xi_4$ increases with area fraction for both $R=1.2$ and $R=1.4$. As with the increase in relaxation time, the effect is greater for $R=1.2$. Lines are a guide to the eye.}
\label{fig:xi4_both}
\end{figure}

\emph{Simulation details.} We investigate the mutual information in computer simulations of two systems. The first is a binary mixture (50:50) of small (radius $\sigma$) and large ($R\sigma$) hard disk particles in 2D. We look at systems with $R=1.4$ and $R=1.2$. In both cases the system exhibits dynamic slowing down (Fig.~\ref{fig:angell}) and other glassy behaviour (e.g. dynamic heterogeneity, Figs.~\ref{fig:xi4_both} and \ref{fig:chi4}), but at $R=1.2$ there is much more crystalline order. We look at systems with area fractions $0.70 \leq \phi \leq 0.80$. Monodisperse hard disks undergo a liquid-hexatic transition at around $\phi=0.71$ and a hexatic-solid transition at around $\phi=0.72$ \cite{bernard:2011}.

The system evolves with Monte Carlo dynamics: a trial move involves shifting a particle to a random position somewhere in an $0.05 \times 0.05 \sigma^2$ square centred on its original position. If the move does not lead to an overlap with any other particle then the move is accepted. We measure time in Monte Carlo sweeps: one sweep involves $N$ attempted moves, where $N$ is the number of particles in the system ($N = 20000$). Periodic boundary conditions are used.

\begin{figure}[htbp]
\includegraphics[height=60mm]{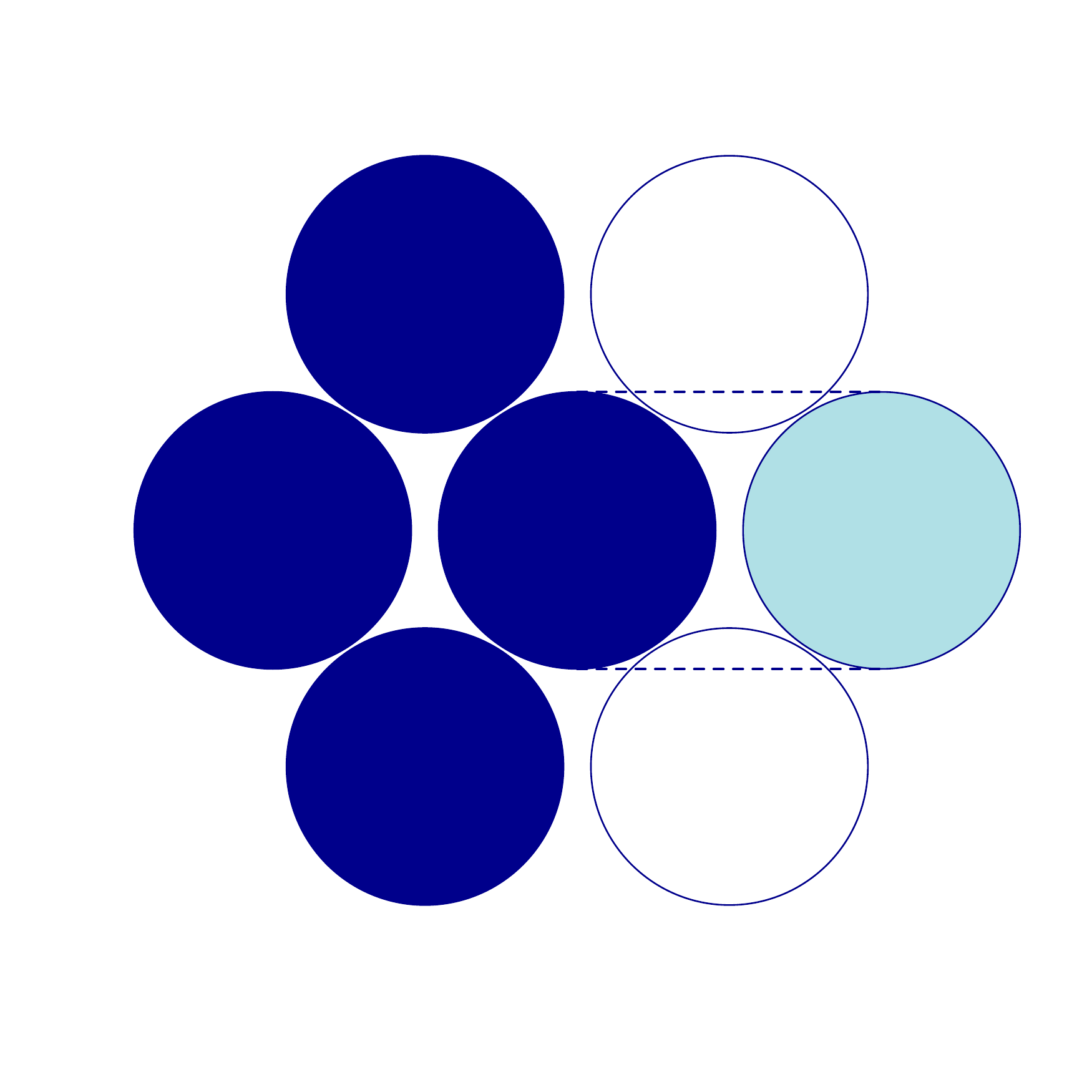}
\caption{The kinetic constraints in the 2-TLG model: for the central particle to move into the right-most vacancy it passes through the mutually neighbouring sites. These must be empty for the move to be accepted. }
\label{fig:tlgrules}
\end{figure}

\begin{figure}[htbp]
\includegraphics[height=60mm]{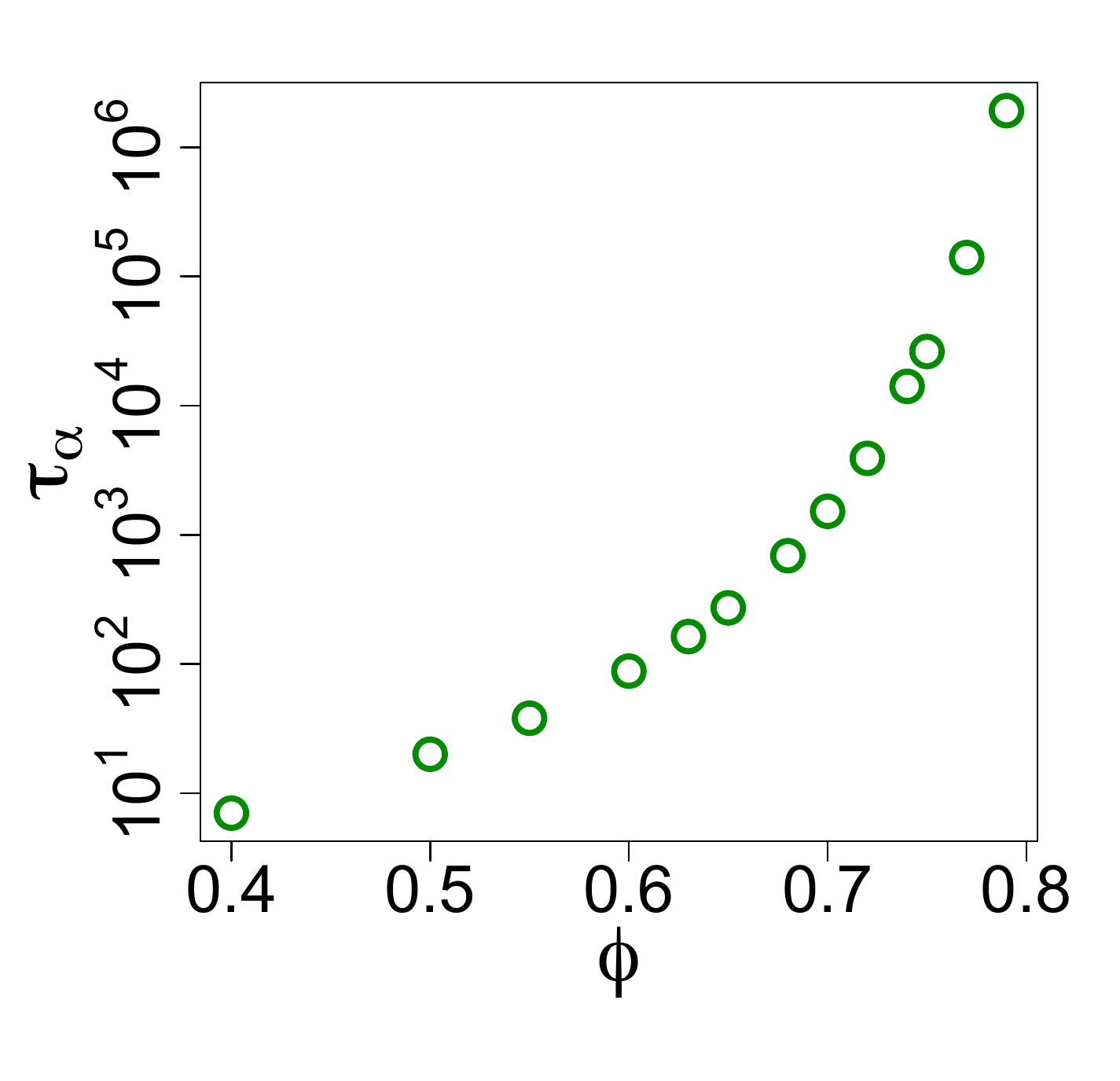}
\caption{Relaxation time against density for the 2-TLG.}
\label{fig:tlgangell}
\end{figure}

The second system is the two-vacancy-assisted-hopping triangular lattice gas (2-TLG). This is a lattice gas model of a glass-forming fluid introduced in \cite{kronig:1994}. Hard particles sit on a two-dimensional triangular lattice.  Monte Carlo dynamics are used to evolve the system: a random particle is chosen and an attempt made to move it to a random neighbouring site. The move is only accepted if the neighbouring site is vacant, and if both sites that are mutual neighbours of the particle's starting site and the trial-move site are also vacant (see Fig.~\ref{fig:tlgrules}). As the system is a lattice gas there is no static structure. However, the (2)-TLG is known to slow down dramatically (see Fig.~\ref{fig:tlgangell}) and become increasingly heterogeneous as its density is increased \cite{pan:2005}. 

\begin{figure}[htbp]
\includegraphics[height=40mm]{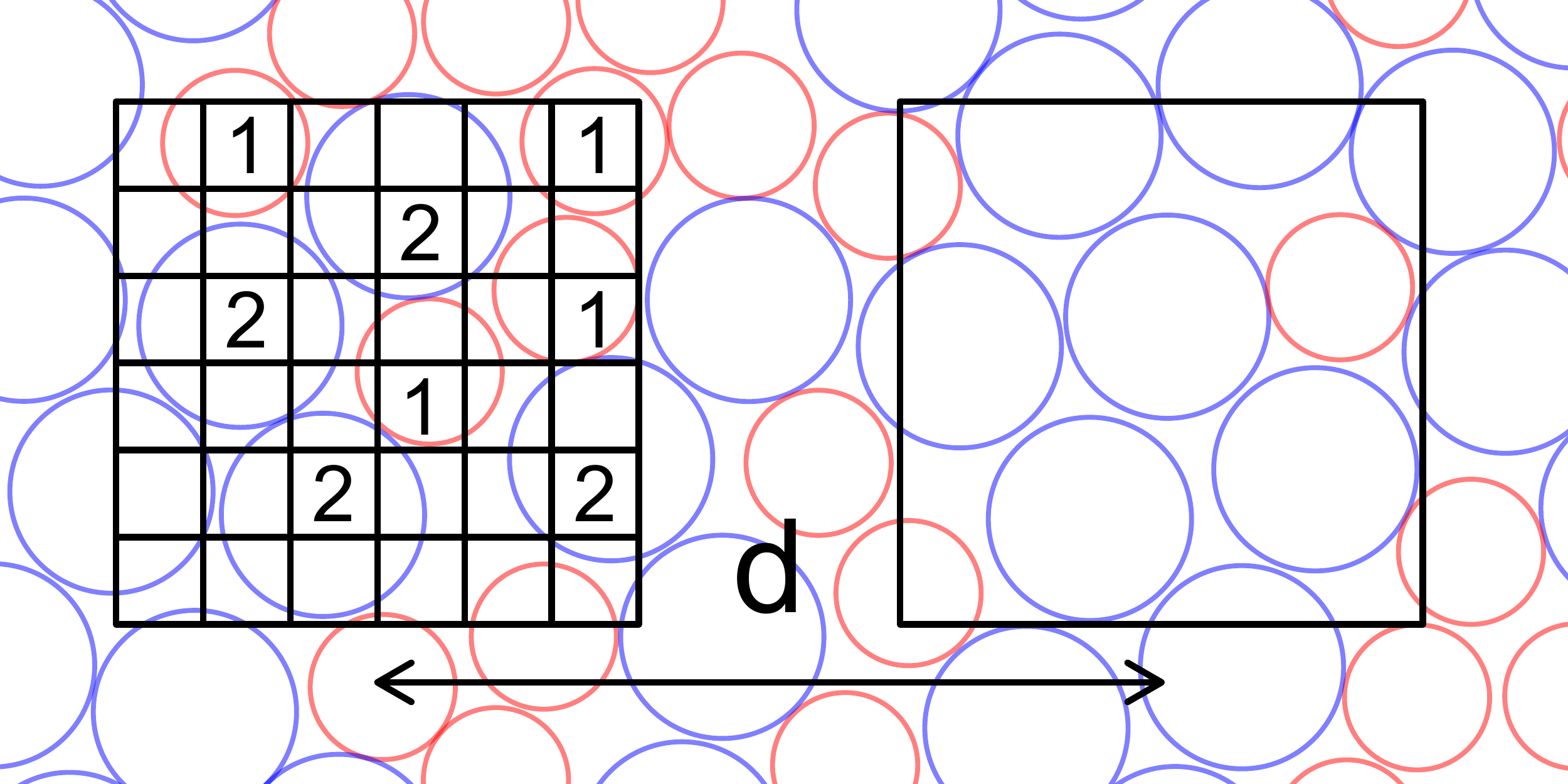}
\caption{The system for discretising the patch configuration in the MC hard disk system (not to scale). A grid is superimposed on the patch and pixels that include a particle centre are marked 1 or 2 depending on particle type. Empty space is encoded with zero. The patches are represented by a 3-ary number. Mutual information is measured between two patches separated by a distance $d \sigma$.}
\label{fig:MC2D_patches}
\end{figure}

\emph{Mutual information based measurements of structure.} In the binary disk system we take $1.8 \times 1.8 \sigma^2$ square patches (each patch is large enough to hold up to six small particles). The distance between the patch centres is $d$ as shown in Fig.~\ref{fig:MC2D_patches}. We discretise the patches onto a $6 \times 6$ grid and represent each configuration as a 3-ary number (0 - empty space; 1 - small particle; 2 - large particle). We assume the joint probability distribution of two patches separated by a given distance (note that the displacement is always along the axis of the square) is stationary over space. As such, we sample the probability distribution over space and multiple instances of the system: the positions and orientations of the sampled pairs of patches are chosen at random.

For the TLG there is no need to discretise the patches. We use hexagonal patches (of varying side length) and encode them as binary numbers (as there is only one particle type).

Mutual information is calculated directly from the histogram of patch values. It is possible to calculate mutual information from continuous distributions \cite{kraskov:2004} but the nature of the patches - they contain an unknown number of either type of particle and therefore do not have a set dimensionality - makes this difficult. These issues do not affect the discretised patches, which always consist of $n^2$ symbols. By the information-processing inequality \cite{cover:1991} the mutual information between discretised patches is a lower-bound on the mutual information between continuous patches. 

The histogram method for estimating mutual information has a positive systematic bias. Finite-sample corrections exist \cite{grassberger:1988} \cite{roulston:1999} although we obtained better results by measuring entropies at different sample numbers and fitting curves to estimate errors (see appendix~\ref{apx:errors}). Equation~\ref{eqn:weuse} is used to calculate mutual information from the entropy measurements. The mutual information static length ($\xi_\mathrm{mi}$) is defined as the first moment of the distribution of patch mutual information with separation, $d$. 

\begin{equation}
\xi_\mathrm{mi} = \frac{\sum_d d \times I(X;Y_d)}{\sum_d I(X;Y_d)} \label{eqn:moment}
\end{equation}

When $d < 1.8 \sigma$ some of the mutual information measured is due to the overlap between the two patches. To avoid this we measure the mutual information between one full-size patch and the non-overlapping part of the other patch. We normalise the mutual information by its theoretical maximum (the entropy of the larger patch) to give a value that is comparable at all $d$.

\emph{Conventional measurements of structure.} We wish to compare any static structure we might find to the length-scale of the dynamical heterogeneity in the system. To do so, we calculate the dynamic length using a four-point correlation function approach similar to that in \cite{lacevic:2003} which is described in appendix~\ref{apx:dynamics}.

\begin{figure*}[htbp]
\includegraphics[height=60mm]{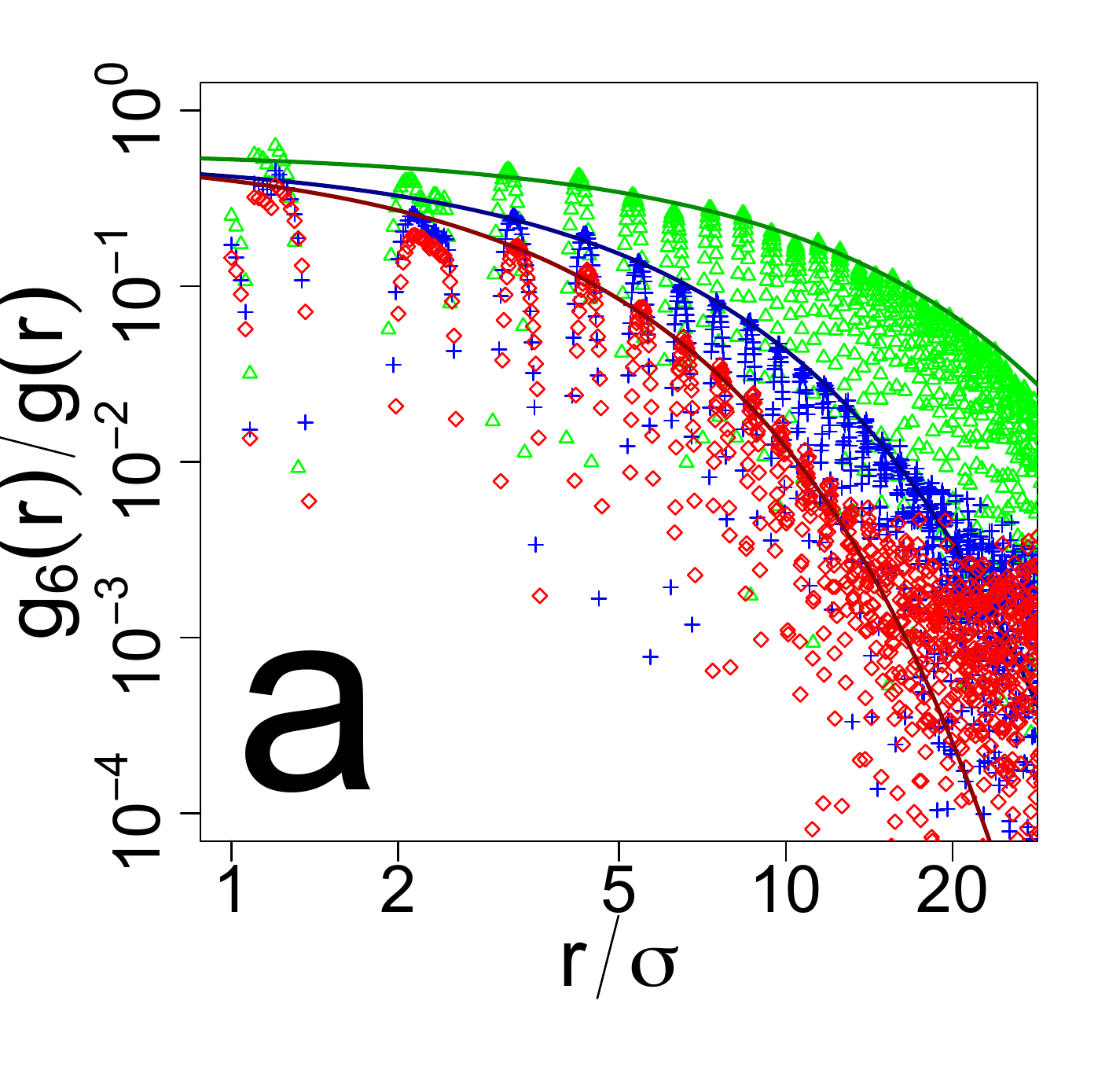}
\includegraphics[height=60mm]{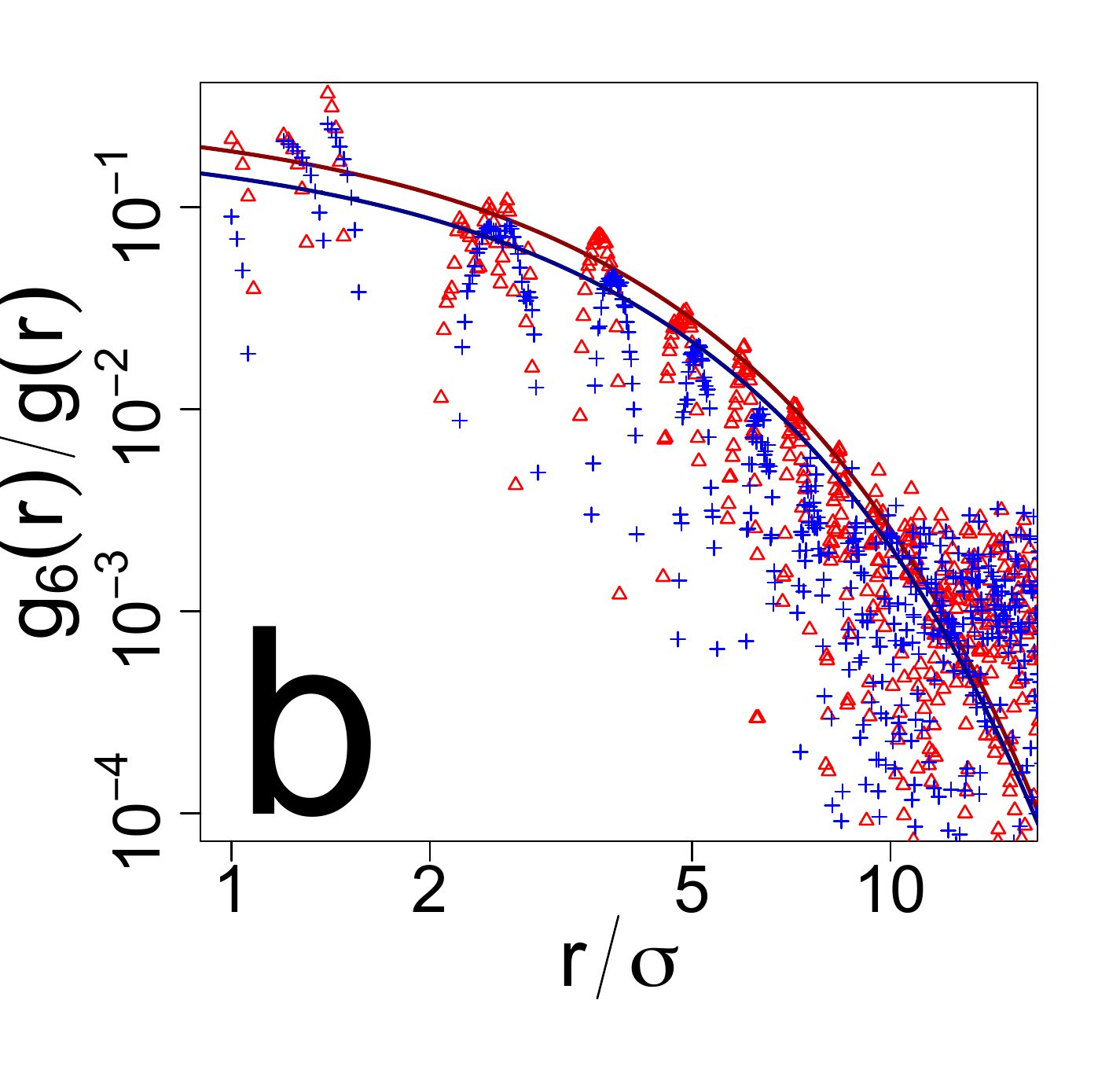}
\includegraphics[height=60mm]{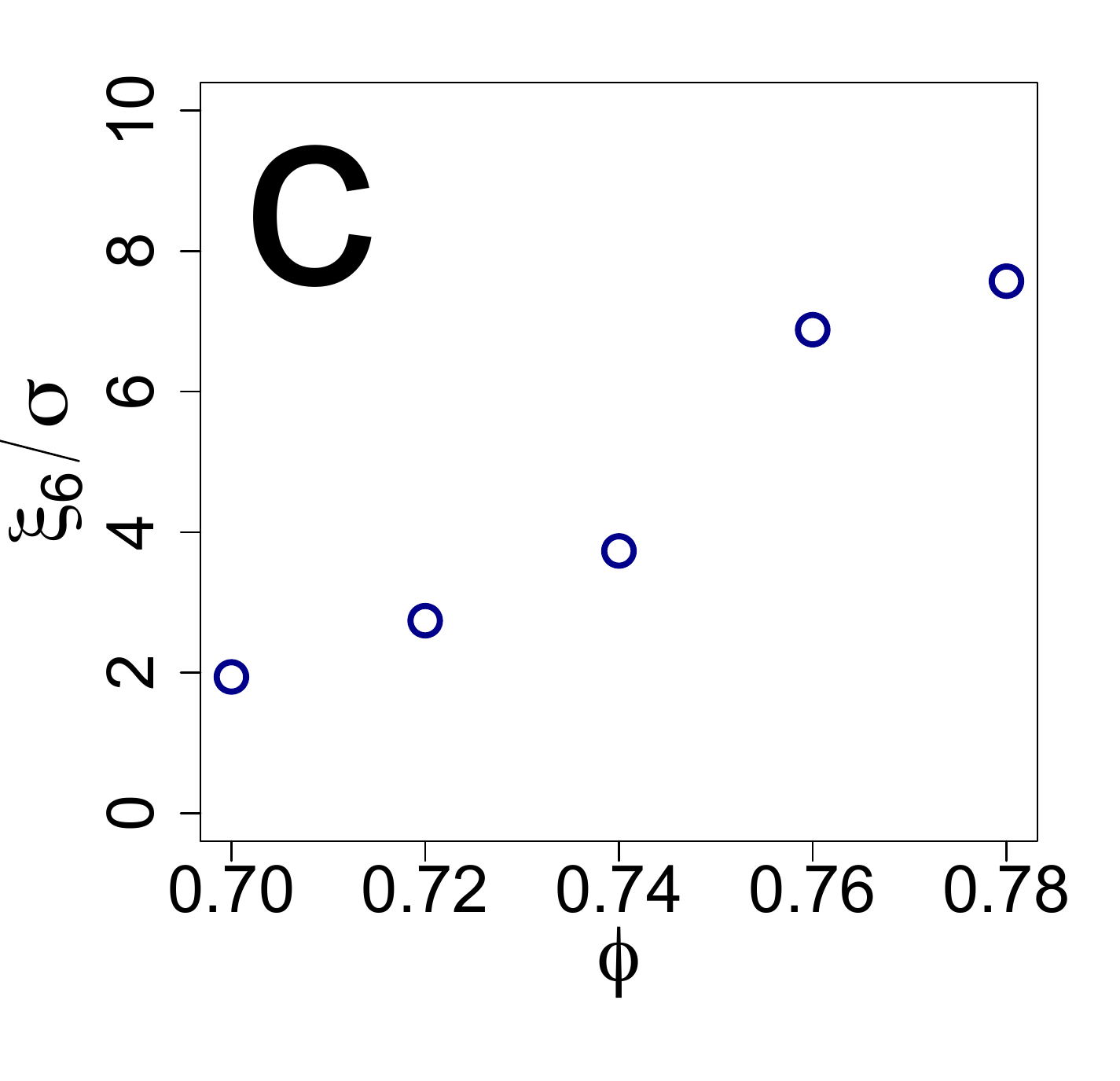}
\includegraphics[height=60mm]{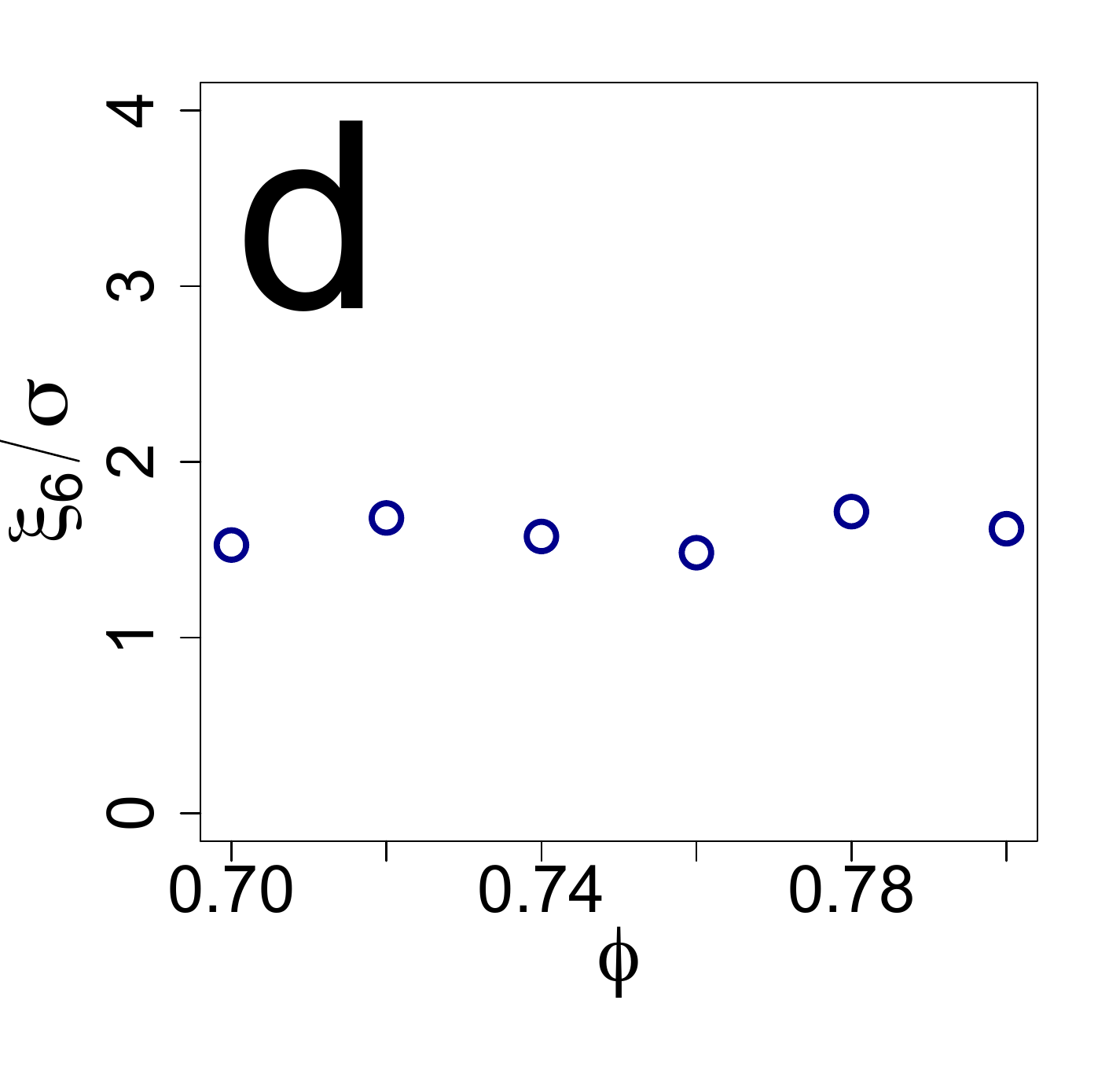}
\caption{Ornstein-Zernike fits and lengths of $\Psi^6$ for $R=1.2$ (a,c) and $R=1.4$ (b,d). The lower plots show the extracted lengths ($\xi_6$) against $\phi$. There is a clear increase in the length of hexatic order for the $R=1.2$ system (c) that is absent when $R=1.4$ (d).}
\label{fig:psi6fits}
\end{figure*}

We also compare the mutual information lengths to other order parameters in the binary hard disk systems: the hexatic order parameter  $\Psi^6_j = \sum_{k \in n(j)} \exp [i6\theta_{jk}]$ ($n(j)$ are the neighbours of particle $j$; $\theta_{jk}$ is the angle between particles $j$ and $k$) \cite{hamanaka:2007}; and local $s_2$ \cite{tanaka:2010}. In both cases a length-scale is extracted by fitting a 2D Ornstein-Zernike envelope \cite{tanaka:2010} to the normalised correlation function of the order parameter (see Fig.~\ref{fig:psi6fits}). $g_x(r)$ is the correlation function and $\xi_x$ the length-scale of a particular order parameter: here we use $g_6(r)$ and $g_{s^2}(r)$ for the correlation functions of $\Psi^6$ and local s$^2$ respectively.
\begin{equation}
 \frac{g_{x}(r)}{g(r)} \propto  r^{-1/4}\exp( - r / \xi_x) 
\end{equation}

Local $s^2$ is calculated from the individual pair correlation functions $g_i^a(r)$ and $g_i^b(r)$

\begin{equation}
g_i^a(r) = \Big < \frac{1}{2 \pi r \Delta r \rho (N-1)} \sum_{j \in A} \delta(r-r_{ij}) \Big >_{10 \tau_\alpha}
\end{equation}
\begin{equation}
g_i^b(r) = \Big < \frac{1}{2 \pi r \Delta r \rho (N-1)} \sum_{j \in B} \delta(r-r_{ij}) \Big >_{10 \tau_\alpha}
\end{equation}

$A$($B$) is the set of small (large) particles. $N$ is the number of particles in the system and $\rho$ is the system density. 

We average over $10 \tau_\alpha$ to remove short-term fluctuations and to ensure that $g(r)$ is adequately sampled. The final particular $s^2$ is the sum of contributions from both correlation functions:
\begin{equation}
s^2_i = -\frac{\rho}{2} \sum_{k \in \{a,b\}} \int_o^{r_k^*}dr \Big [ g_i^k(r) \ln g_i^k(r)  - (g_i^k(r) -1)   \Big]
\end{equation}

The integration of $s^2$ should ideally be between zero and $+\infty$. Practically, it is cut off at a value that is large enough to take in many shells of surrounding particles. This may be slightly different to the method implemented in \cite{tanaka:2010} but we believe out method should capture any static length-scale.

\begin{figure*}[htbp]
\includegraphics[height=55mm]{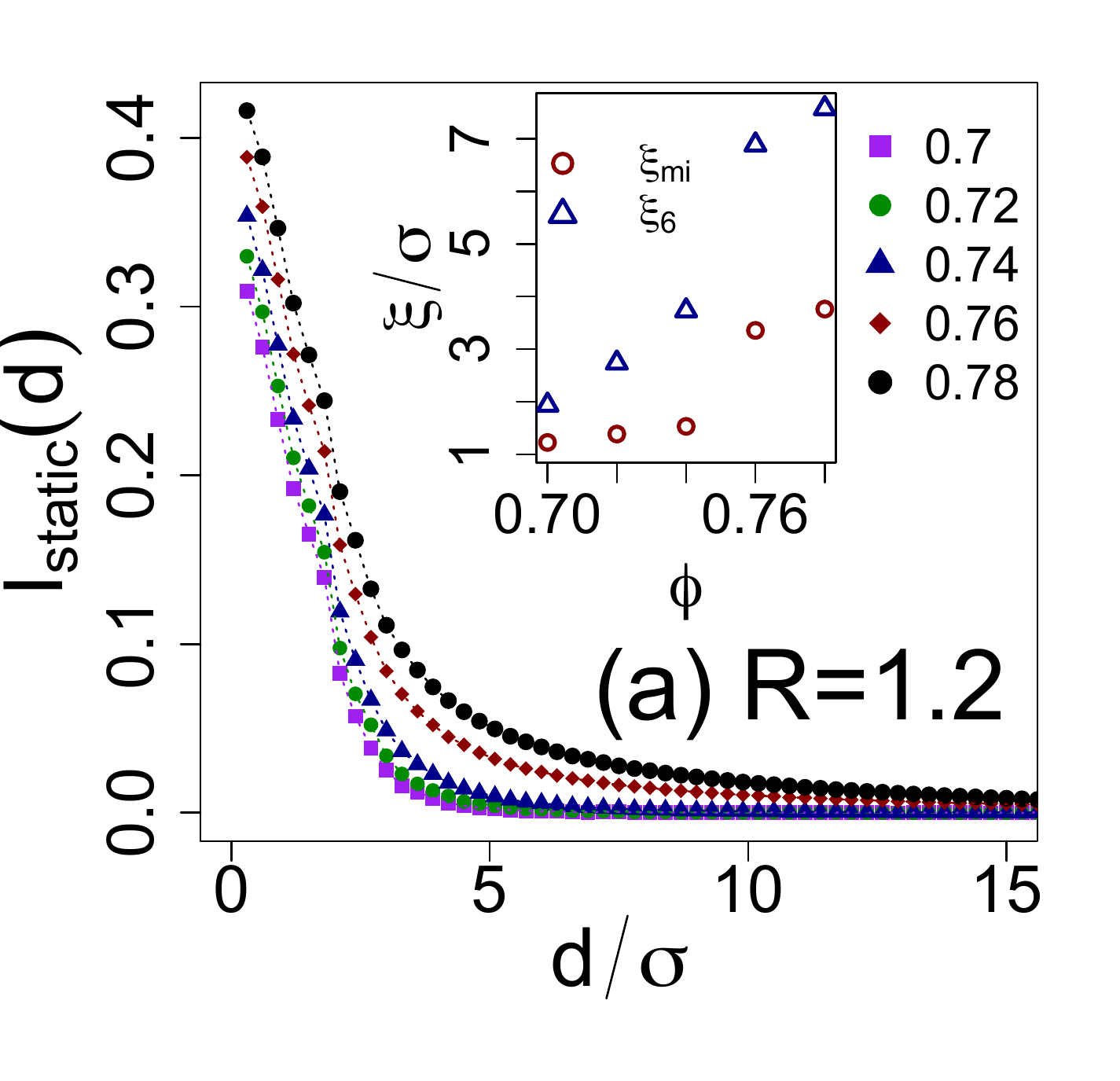}
\includegraphics[height=55mm]{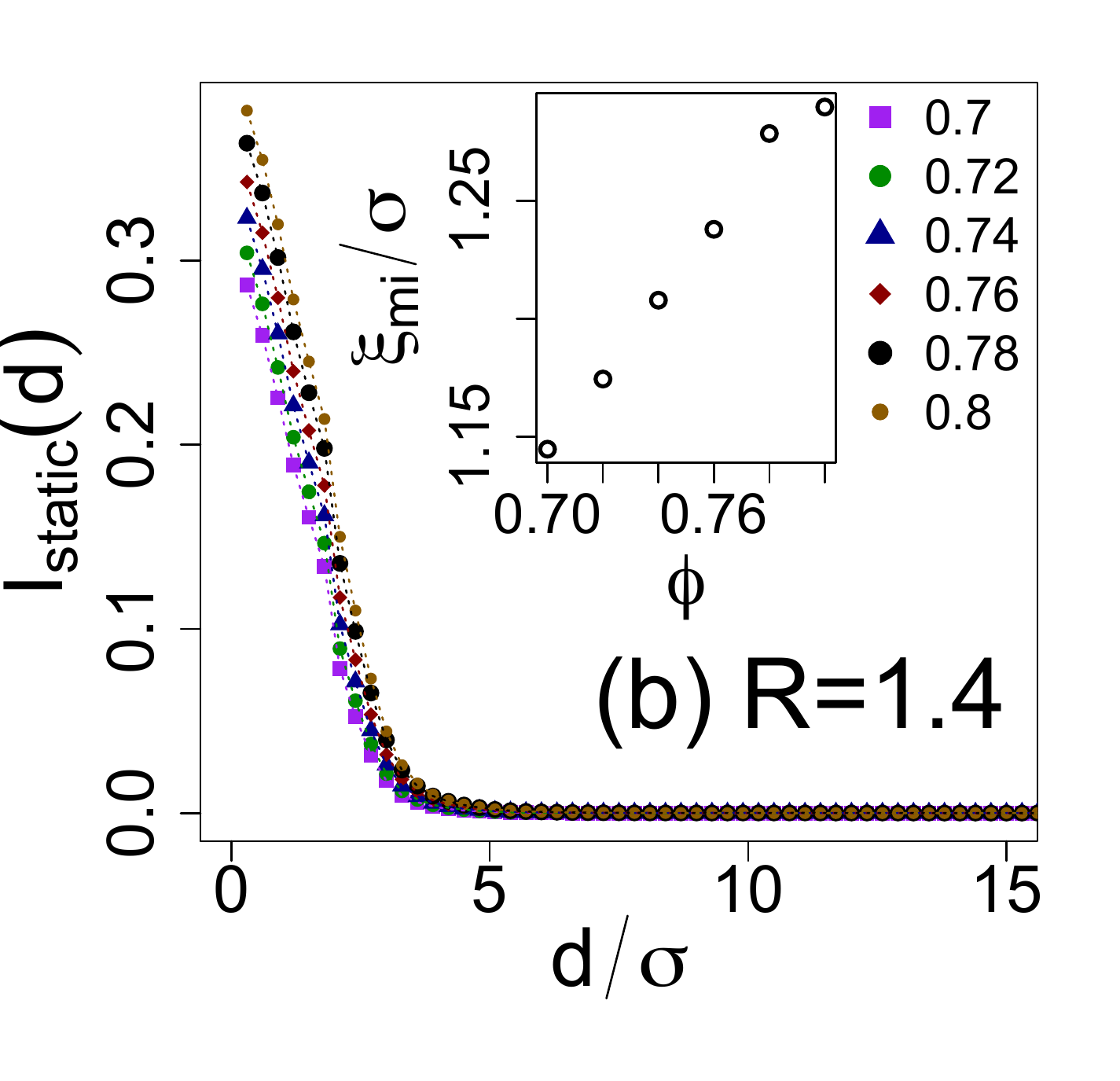}
\includegraphics[height=55mm]{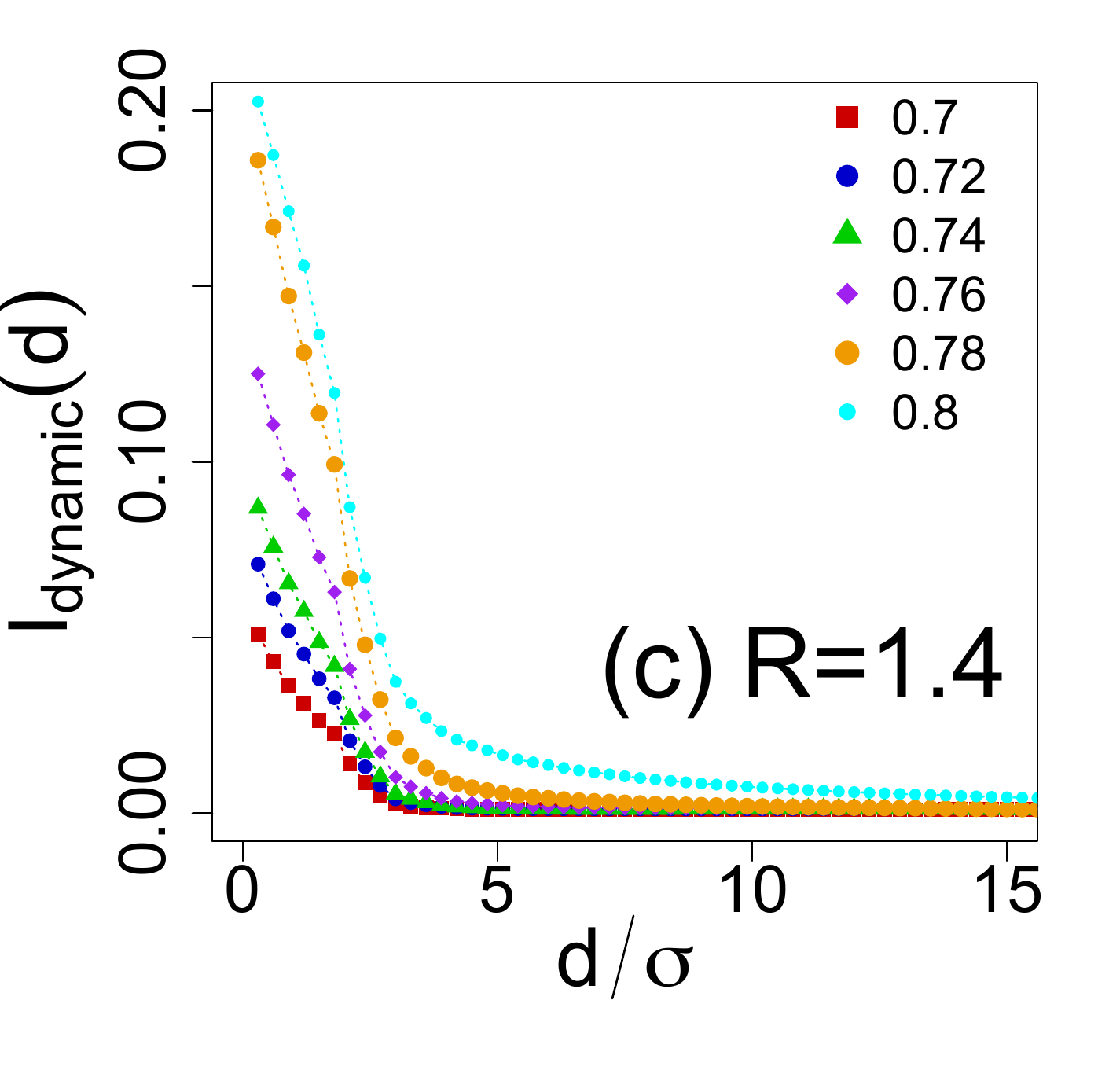}
\caption{Plots of the mutual information between patches ($I(d)$) against patch separation $d$ for the binary hard disk system with $R=1.2$ and $R=1.4$. The values have been corrected for overlapping patches and normalised so that at each distance the maximum possible mutual information is one. The left and central plots are of mutual information between patches encoding static information. The right plot shows mutual infomration betwen patches encoding dynamic information for the $R=1.4$ system. The (static) mutual information lengths (and the $\Psi^6$ length for $R=1.2$) are plotted in the insets.}
\label{fig:mi} \label{fig:midyn}
\end{figure*}

\section {Results}

\subsection {Binary hard disks}

As $\phi$ is increased from 0.7 the relaxation times in both systems increase significantly (Fig.~\ref{fig:angell}) with the $R=1.2$ system slowing more. The relaxation times, $\tau_\alpha$, are measured from self intermediate scattering functions. Vogel-Fulcher-Tamman fits ($\tau_\alpha = \tau_0 \exp [D \phi / (\phi_0 - \phi$)]) are used to obtain $\phi_0$, the ideal glass transition packing fraction. This is used later when comparing length-scales.

Both systems also become dynamically heterogeneous. Plots of $\chi_4$ (Fig.~\ref{fig:chi4}) indicate that both systems are dynamically heterogeneous over intermediate times and that the maximum heterogeneity increases as the system becomes more dense. Figure~\ref{fig:xi4_both} shows the dynamical correlation lengths calculated at the time of maximum $\chi_4$ for each $\phi$. These show the range of the dynamic correlations increasing with $\phi$. Again, this is more pronounced in the $R=1.2$ system. The details of how $\chi_4$ and $\xi_4$ were calculated are in appendix~\ref{apx:dynamics}.

We begin by considering the distribution of mutual information with patch separation distance (Fig.~\ref{fig:mi}). There is little change in the distribution with $\phi$ for the $R=1.4$ system: the mutual information increases slightly at short distances and there is a small but consistent increase in the mutual information length as the density is increased. This is in marked contrast to the $R=1.2$ system (as is be expected given its hexatic ordering). In this case the mutual information decays much more slowly with distance at the higher density state points. The increase in mutual information length seems consistent with the increase in $\xi_6$.

\begin{figure*}[htbp]
\includegraphics[height=70mm]{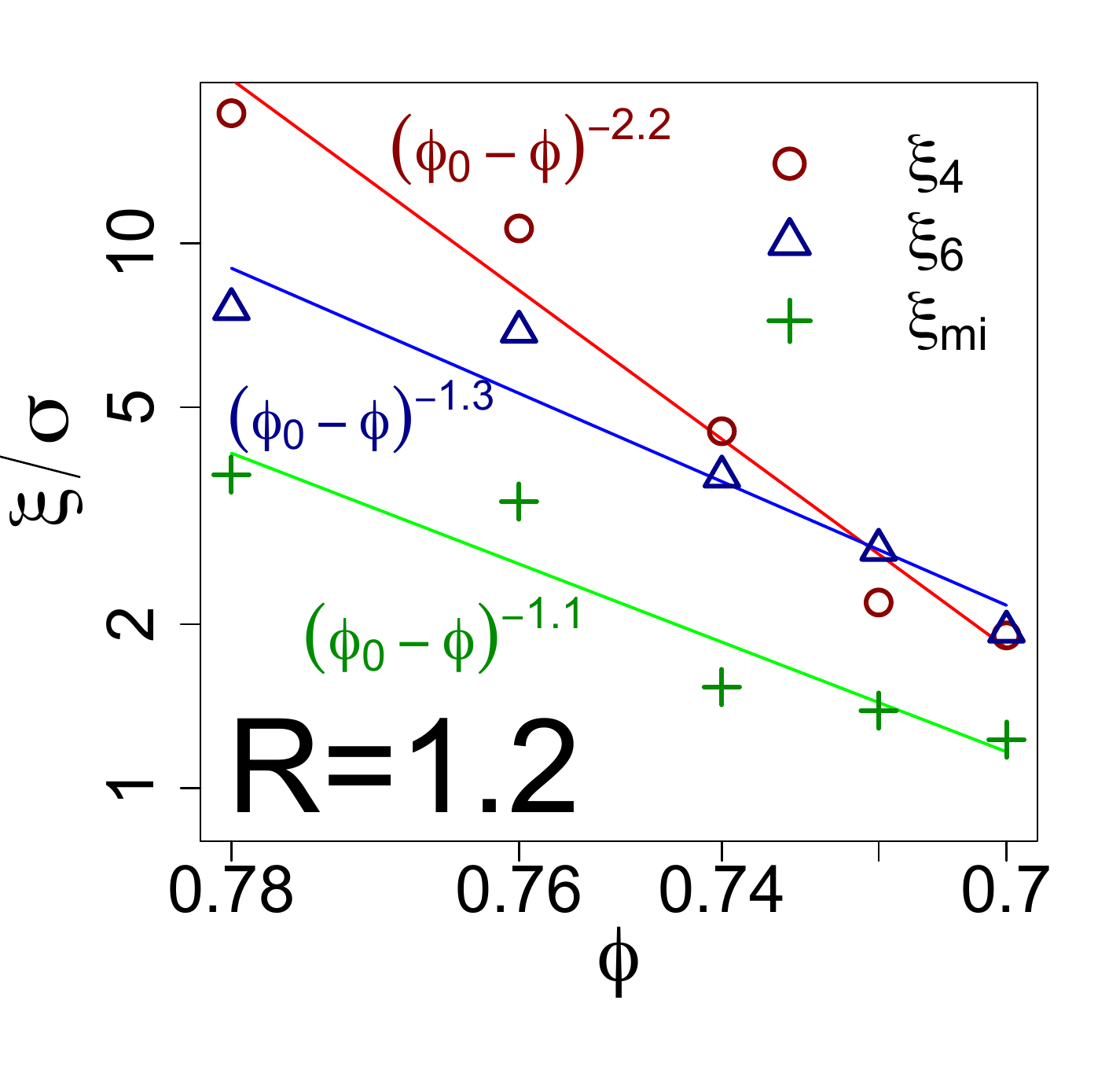}
\includegraphics[height=70mm]{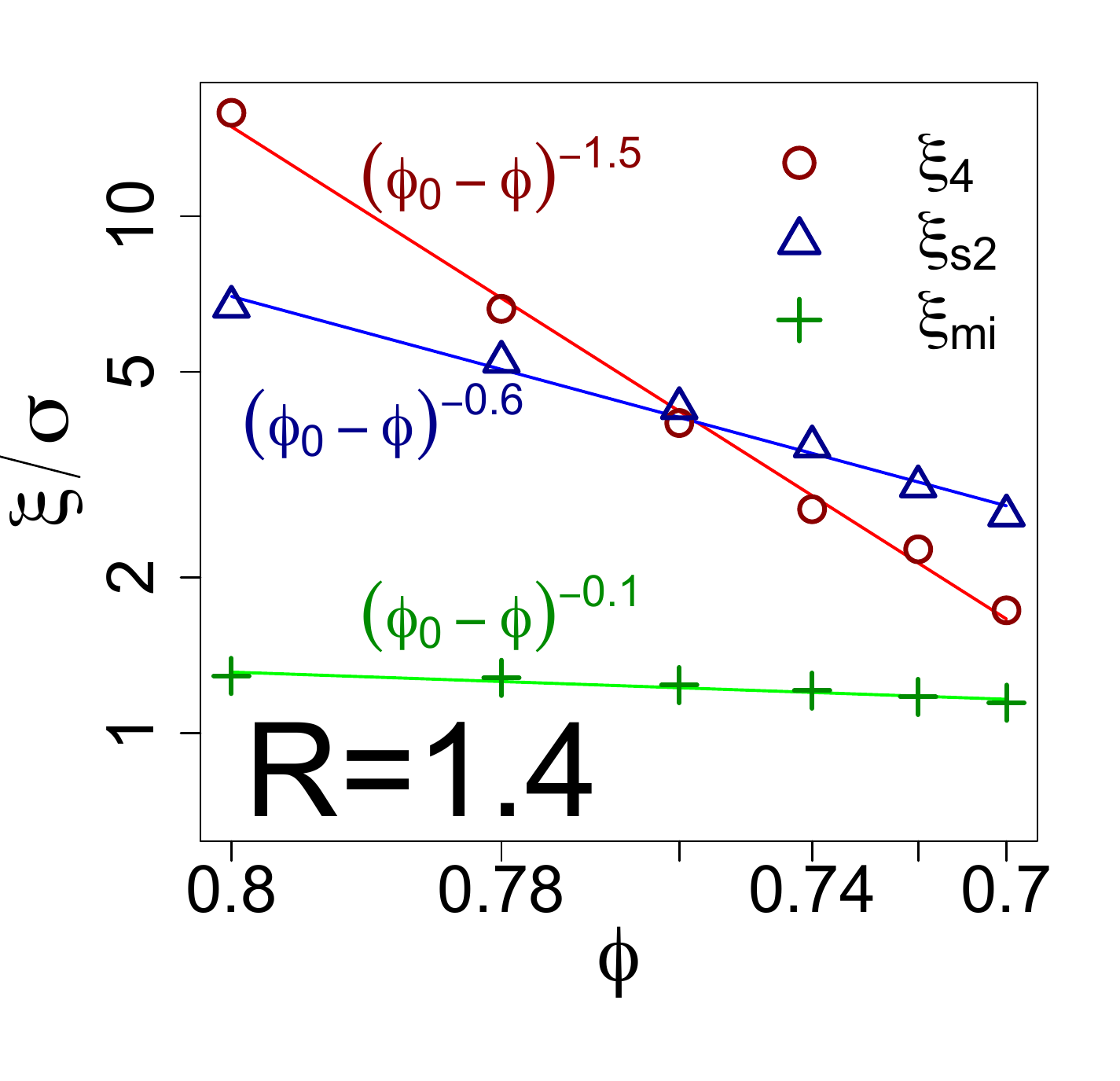}
\caption{A comparison of lengths in the binary hard disk $R=1.2$ and $R=1.4$ systems. The lines are fits of $\xi \propto (\phi_0-\phi)^b$ where $\phi_0$ is taken from the VFT fit of $\tau_\alpha$ for the system. $\xi_4$ is the dynamical correlation length; $\xi_\mathrm{mi}$ is the (static) mutual information length; and $\xi_6$ and $\xi_{s^2}$ are the correlation lengths of $\Psi^6$ and local s$^2$ respectively. The exponent for $\xi_\mathrm{mi}$ in the $R=1.4$ system is $b = -0.1$ with a standard error of 0.01: as density is increased the growth in $\xi_\mathrm{mi}$ is small, but non-zero.}
\label{fig:milength}
\end{figure*}

This is apparent in Fig.~\ref{fig:milength} where the various lengths are plotted against the reduced area fraction, $\phi_0 - \phi$. The mutual information and $\Psi^6$ lengths ($\xi_\mathrm{mi}$ and $\xi_6$) increase at similar rates in the $R=1.2$ system. However neither increases as fast as the dynamical heterogeneity length, $\xi_4$. The difference in static and dynamic length-scales is more pronounced in the $R=1.4$ system. Here the mutual information length barely increases at all whereas the dynamic length at $\phi=0.8$ is an order of magnitude greater than at $\phi=0.7$.

As the hexatic order parameter varies little for the $R=1.4$ system we measure the local $s^2$ length. The correlations in this quantity have been used to detect order in various glass-forming liquids \cite{tanaka:2010}. It measures (in its global form) the pairwise contribution to extra configurational entropy of the system compared to an ideal gas \cite{truskett:2000}. Unlike the static mutual information length, $\xi_{s^2}$ increases with system density (although not as much as $\xi_4$). 

Essentially, $s^2$ measures the peakedness of a pair correlation function. It is averaged, in this case, over a trajectory of the system: so a particle that moves little and is surrounded by similar particles will have a rather spiked $g(r)$; particles that move around a lot will smooth out their $g(r)$. The peakedness of individual $g(r)$ that are averaged in this way will be sensitive to dynamic heterogeneity. If we assume mutual information measures \emph{any} true increase in structural correlation length, it is uncertain whether increasing $\xi_{s^2}$ is measuring an increasing static length or is confounded by the increasing dynamic length-scale. The fact that the mutual information length shows no such increase suggests that the second possibility is likely. 

Finally, we look at the dynamic mutual information between patches. This is calculated similarly to the static mutual information, but we no longer encode the particle type and instead encode the mobility of the particle. The mobility is taken from the overlap function (Equation~\ref{eqn:overlap}) used to calculate the four-point correlation functions. The patch is constructed such that mobile particles are encoded with a one and immobile particles and empty space are encoded with zero. The information the patches contain is the position of mobile particles in the system. Unlike the static mutual information, the dynamic mutual information shows a notable increase at short distances and extends to longer distances as $\phi$ is increased (Fig.~\ref{fig:midyn}).

\subsection {Triangular Lattice Gas}
\label{sec:tlg}

\begin{figure}[htbp]
\includegraphics[height=60mm]{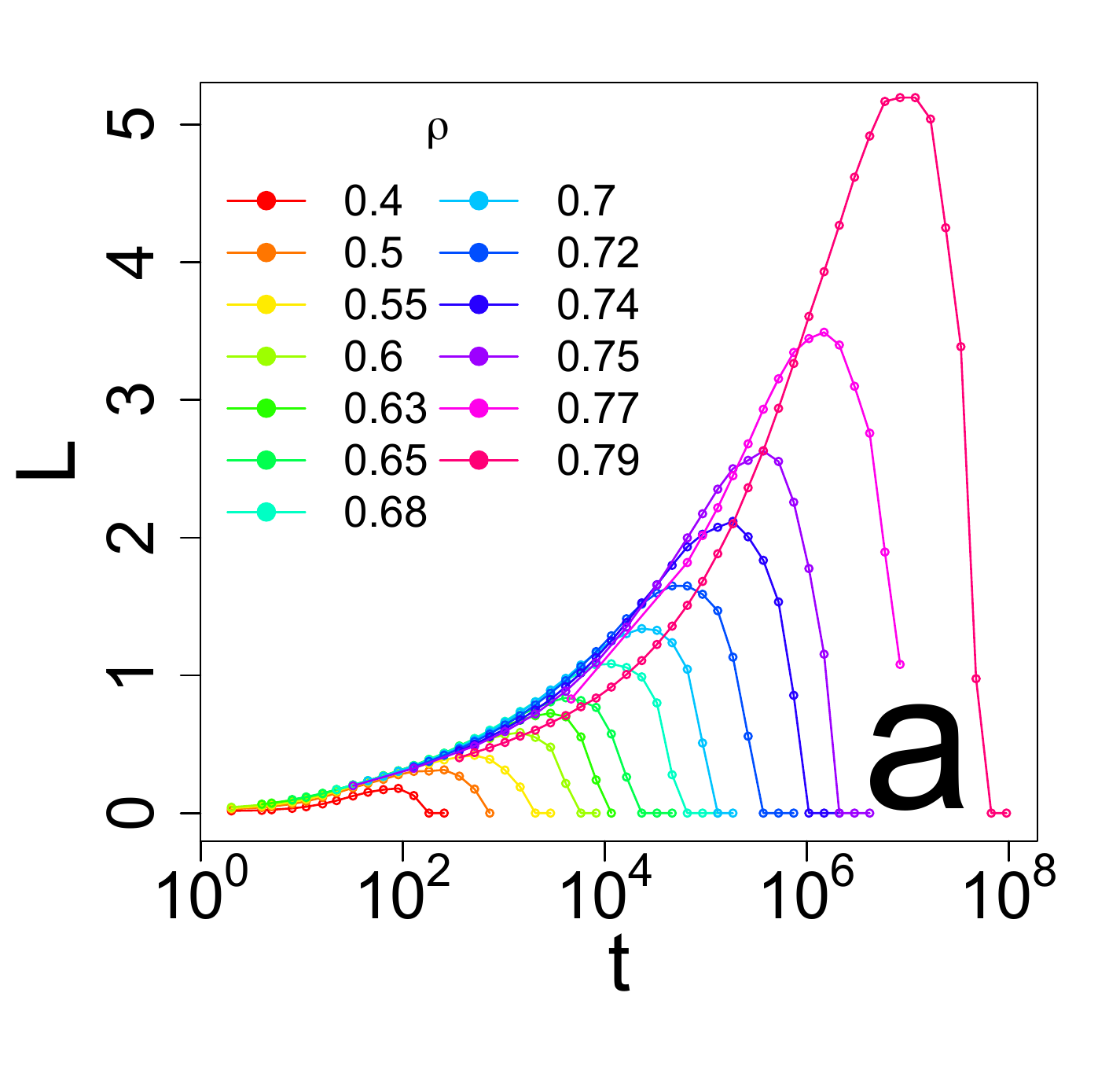}
\includegraphics[height=60mm]{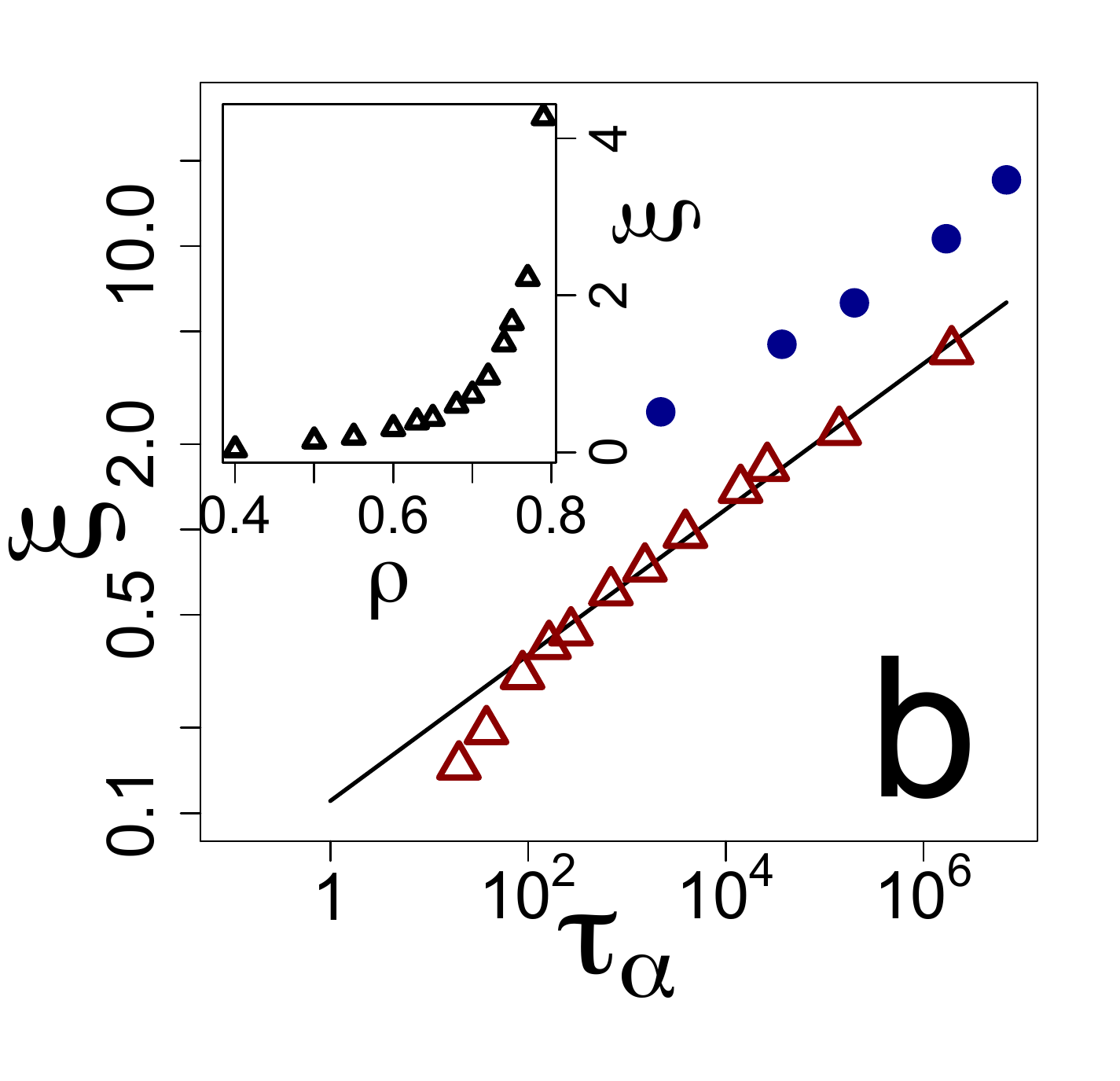}
\caption{(a) the dynamic mutual information length ($\xi_\mathrm{dyn, mi}(t)$) for the 2-TLG at different densities and with mobility measured over different times, $t$; (b) the dynamic mutual information length at $t=\tau_\alpha$ for each density ($\xi_\mathrm{dyn, mi}(\tau_\alpha)$ - red triangles). The blue circles are dynamic heterogeneity length data from \cite{pan:2005} for the 2-TLG. For higher densities (higher $\tau_\alpha$) the lengths scale similarly. The inset shows $\xi_\mathrm{dyn, mi}(\tau_\alpha)$ plotted against density.}
\label{fig:tlglength}
\end{figure}

The 2-TLG, being a kinetically constrained model, has no static structure by design. As expected, no static structure was found using patch mutual information. We calculated the dynamic mutual information (as described above) using patches of various radii. Here we focus on the lengths obtained from radius one (i.e. point) patches. These contain no extra structural information and so are directly comparable to the dynamic lengths obtained from four-point correlation functions. The dynamic mutual information lengths are obtained as before (Equation~\ref{eqn:moment}). The mobility of particles is again measured by an overlap function: if the particle has moved from its original position after time $t$ then it is mobile. 

Figure~\ref{fig:tlglength} shows the dynamic mutual information length, $\xi_\mathrm{dyn, mi}(t)$, measured at different $t$ for various system densities ($\rho$). It is clear that dynamical heterogeneity is an intermediate time-scale phenomenon: each curve peaks at a time proportional to the relevant relaxation time ($\tau_\alpha$). As expected, this maximum length increases as the system becomes more dense. Figure~\ref{fig:tlglength} also shows the dynamic mutual information lengths at $\tau_\alpha$, $\xi_\mathrm{dyn, mi}(\tau_\alpha)$, compared to existing four-point correlation measurements (from \cite{pan:2005}). The scaling of length with relaxation time with exponent 1.4 agrees well for both sets of measurements.

\section {Discussion and Conclusions}

In this paper, we have shown that the mutual information between patches can be used to measure static order in amorphous systems. It is possible to extract a length-scale from such measurements, and in the system where we had direct access to a relevant order parameter the mutual information length was in agreement. The dynamic mutual information measurements are useful in that they show the difference in the growth of correlations in the $R=1.4$ hard disk system as $\phi$ increases. However, it is not an optimal way to measure dynamic heterogeneity: whereas the 4-point-correlation function \cite{lacevic:2003} depends only on the mobility of pairs of particles at some distance, the dynamic mutual information contains unneeded information of the configuration of mobile particles within the patches. In the case where the patches are reduced to point size (as for the 2-TLG, Section~\ref{sec:tlg}) this issue is removed and we recover the behaviour of the correlation function.

Our results support the conclusion that there need not be a growing static length coupled with the dynamics of glassy systems, agreeing with previous work such as \cite{charbonneau:2012,malins:2012,berthier:2007}. The mutual information length changes little with $\phi$ for the $R=1.4$ hard disk system despite the vast increase in relaxation time and dynamical length over the same range. In the $R=1.2$ system both the hexatic order parameter and the mutual information show a length that increases significantly with $\phi$: however, the dynamic length increases markedly faster. It should be noted that in the $R=1.4$ system the mutual information length casts some doubt on the interpretation of $\xi_{s^2}$, the local s$^2$ length, and that $\xi_{s^2}$ does not increase as quickly as $\xi_4$, the dynamic length, as we increase $\phi$. It should be noted that we did have trouble measuring s$^2$ exactly as specified in \cite{tanaka:2010} and so this may have influenced our results.

These conclusions are compatible with dynamical facilitation \cite{keys:2011} or dynamical phase-transition \cite{garrahan:2007} explanation of the glass transition, rather than scenarios which invoke strongly growing static order such as Random First Order Theory \cite{lubchenko:2007}. It may be relevant that the hard disk system investigated here has purely repulsive interactions. There is reason to suppose that local structure is more important in systems with attractive potentials \cite{coslovich:2011}. It is possible that the mutual information between patches would behave differently in such systems.  

\begin{acknowledgments} 
We thank R. Jack and P. Charbonneau for helpful discussions. A.J.D. is funded by EPSRC grant code EP/E501214/1. C.P.R. gratefully acknowledges the Royal Society for financial support. This work was carried out using the computational facilities of the Advanced Computing Research Centre, University of Bristol - http://www.bris.ac.uk/acrc/.
\end{acknowledgments}

\appendix

\section{Compensating for finite sample errors}
\label{apx:errors}

\begin{figure}[htbp]
\includegraphics[height=60mm]{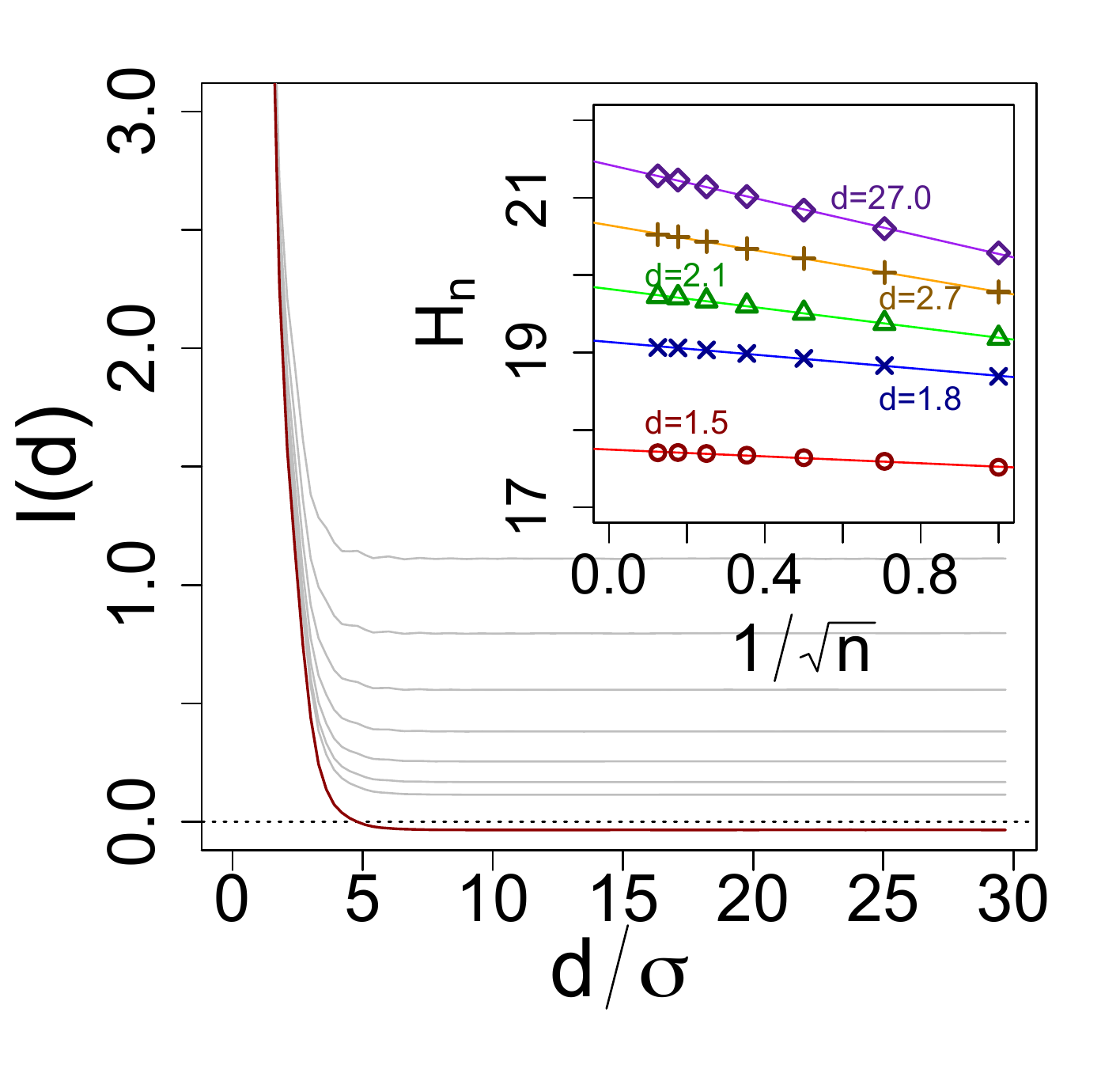}
\caption{The grey lines in the main plot are mutual information estimates (without overlap corrections) with different sample sizes: the number of samples doubles each time from top to bottom. The red line is the estimated true entropy, $H_\infty(X,Y_d)$. The inset shows the first-order error fits that were used to obtain $H_\infty(X,Y_d)$ for various values of $d$.}
\label{fig:misamples}
\end{figure}

We estimate each probability distribution with the frequency distribution obtained by sampling. As we have only a finite number of samples we may not encounter some low probability patch configurations and therefore our estimate of the support of the probability distribution will be too small. Also, the frequencies we measure will fluctuate from their true values which will have the effect of making the estimated distribution less uniform than the true distribution. Both of these effects cause a systematic underestimation in entropy. The effect increases with the size of the probability space (holding the number of samples constant) and so when estimating mutual information using Equation~\ref{eqn:weuse} it is the negative $H(X,Y)$ term that dominates the error. As such, the mutual information will have a positive systematic bias. Figure~\ref{fig:misamples} shows this effect: there is positive mutual information at long distances when we would expect none. This effect reduces as the sample size is increased. 

To estimate the true entropy, $\lim_{d\to \infty} H(X,Y_d)$ for two patches separated by a distance $d$ we measure the entropy at various sample sizes: $H_n(X,Y_d)$ where $n$ is proportional to the number of samples. We assume that the difference between the true and finite sample entropies is given by a series of terms \cite{grassberger:1988}:   
\begin{equation}
H_n = H_\infty + k_1 \Big (\frac{1}{n^b}\Big ) + k_2 \Big ( \frac{1}{n^b} \Big )^2 + \dots
\end{equation}  
By fitting our data to this form we can estimate $H_\infty$.

\begin{figure}[htbp]
\includegraphics[height=60mm]{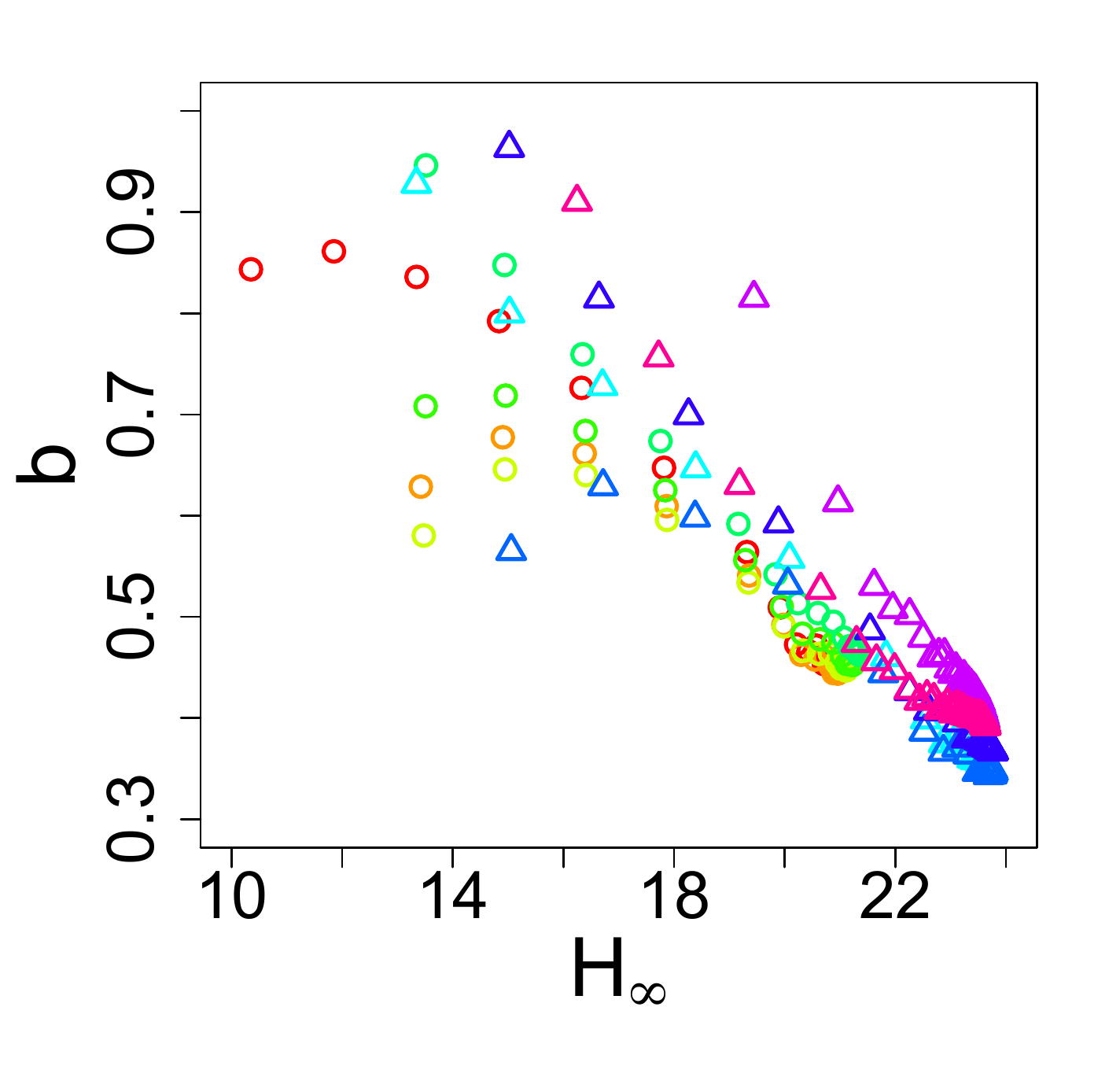}
\caption{The exponent in the error terms versus the estimated true entropy. The different coloured points represent systems with different $\phi$. Triangles are $R=1.2$ systems; circles are $R=1.4$ systems}
\label{fig:bscale}
\end{figure}

Figure~\ref{fig:misamples} shows such a fit using only the first order error term (this technique was used for the $R=1.4$ systems). In this case $b=0.5$ gives a good fit for all non-overlapping patches. The exact value of $b$ varies between the systems and decreases as $H$ increases (Fig.~\ref{fig:bscale}). 

By ignoring higher order terms we overestimate the error and get (unphysical) negative mutual information values. To compensate for this we shift the curve so that the baseline is zero before measuring the mutual information length. Adding the second order term decreases this error although it makes no significant difference to the mutual information length.

\section{Dynamic correlation measurements}
\label{apx:dynamics}

We derive the dynamic correlation length using a four-point correlation function approach similar to that in \cite{lacevic:2003}. To start with we calculate an overlap function for each particle.
\begin{equation}
w_i(t) = \left\{ \begin{array}{rl}
 1 &\mbox{ if $|\mathbf{r}_i(t) - \mathbf{r}_i(0) | > 0.3 \sigma$} \\
  0 &\mbox{ otherwise}
       \end{array} \right. \label{eqn:overlap}
\end{equation}
To find the time span, $\tau_{h}$, over which the system is most dynamically heterogeneous we calculate
\begin{equation}
 \chi_4(t) =   \frac{1}{N \rho} \Big [ \langle Q(t)^2 \rangle - \langle Q(t)\rangle^2 \Big ]
\end{equation}

Where $Q(t) = \sum_i w_i(t)$. The averages are taken over many realisations of the system. For each density we find $\tau_{h}$: the time which maximises $\chi_4(t)$ and use it to calculate a structure factor:
\begin{equation}
S_4(k) = \sum_{ij} (w_i(\tau_h) - \bar{w}(\tau_h))(w_j(\tau_h) - \bar{w}(\tau_h))\exp(ik \Delta r_{ij})
\end{equation}
($k$ is spatial frequency). This is circularly averaged and an Ornstein-Zernike function is fit to the low $k$ part of our data to give $\xi_4$, the dynamic length:
\begin{equation}
S_4(k) = \frac{S_0}{1 + (k \xi_4)^2}
\end{equation}

Figures~\ref{fig:chi4}~and~\ref{fig:xi4_both} show $\chi_4$ and $\xi_4$ against $\phi$. As expected, the maximum heterogeneity is greater and occurs later for higher density systems. The lengths $\xi_4$ grow rapidly as $\phi$ is increased.

\end{document}